\documentclass{article}
\pdfoutput=1
\usepackage{jheppub}
\usepackage{mathrsfs}
\usepackage[utf8]{inputenc}
\usepackage{color,bm,comment,braket}
\usepackage[T1]{fontenc}
\usepackage{ulem}

\newcommand{\del}{\partial}
\newcommand{\beq}{\begin{eqnarray}}
\newcommand{\eeq}{\end{eqnarray}}

\newcommand{\rmd}{\text{d}}
\newcommand{\rme}{\text{e}}

\preprint{RESCEU-11/22, \ KEK-TH-2437}
\title{Formation of Chiral Soliton Lattice}
\author[a,b]{Tetsutaro Higaki}
\author[c]{Kohei Kamada}
\author[d,b]{Kentaro Nishimura}
\affiliation[a]{Department of Physics, Keio University, 3-14-1 Hiyoshi, Yokohama, Kanagawa 223-8522, Japan}
\affiliation[b]{Research and Education Center for Natural Sciences, Keio University, 4-1-1 Hiyoshi, Yokohama, Kanagawa 223-8521, Japan}
\affiliation[c]{Research Center for the Early Universe (RESCEU), Graduate School of Science, The University of Tokyo, Hongo 7-3-1 Bunkyo-ku, Tokyo 113-0033, Japan}
\affiliation[d]{KEK Theory Center, Tsukuba 305-0801, Japan}
\emailAdd{thigaki@rk.phys.keio.ac.jp}
\emailAdd{kohei.kamada@resceu.s.u-tokyo.ac.jp}
\emailAdd{nishiken@post.kek.jp}
\date{June 2022}

\abstract{
The Chiral Soliton Lattice (CSL) is a lattice structure composed of domain walls aligned in parallel at equal intervals, 
which is energetically stable in the presence of a background magnetic field and 
a finite (baryon) chemical potential 
due to the topological term originated from the chiral anomaly. We study its formation 
from the vacuum state, with describing the CSL as a layer of domain-wall disks surrounded by the vortex or string loop, based on the Nambu-Goto-type effective theory. We show that the domain wall nucleates via quantum tunneling when the magnetic field is strong enough.   
We  evaluate its nucleation rate and determine the critical magnetic field strength with which the nucleation rate is no longer
exponentially suppressed. We apply
this analysis 
to the neutral pion in the two-flavor QCD as well as the axion-like particles (ALPs) with a finite (baryon) chemical potential under an external magnetic field.  
In the former case, even 
though the CSL state is more energetically stable than the vacuum state 
and the nucleation rate becomes larger
for sufficiently strong magnetic field,
it cannot be large enough so that
the nucleation of the domain walls is not exponentially suppressed and 
promoted, without suffering from the tachyonic instability of the charged pion fluctuations.
In the latter case, we confirm that the effective interaction of the ALPs generically includes the topological term required for the CSL state to be energetically favored.
We show that the ALP CSL formation is promoted if the magnetic field strength and the chemical potential of the system
is slightly larger than the scale of the axion decay constant.
}

\begin{document}

\maketitle

\section{Introduction}
The domain wall is a 
field configuration connecting two vacua and is identified as
a two-dimensional topological defect. 
If the system has a periodic potential with (infinite) degenerate vacua, which typically appears for the pseudo 
Nambu-Goldstone bosons (pNGBs) associated with the spontaneous breaking of a global U(1) symmetry, it allows
a stack of parallel domain walls, well-known as the chiral soliton lattice (CSL).
This state universally appears from condensed matter physics to high-energy physics.
The CSL-type magnetic structure has been originally studied in chiral magnets~\cite{dzyaloshinskii1964theory} and experimentally observed \cite{togawa2016symmetry} (see ref.~\cite{kishine2015theory} for review).
In high energy physics, it has been shown, based on a low-energy effective theory that takes into account
the chiral anomaly~\cite{Son:2004tq,Son:2007ny},
that the ground state of QCD at finite baryon chemical potential $\mu_{\textrm{B}}$ under a sufficiently strong magnetic field is the CSL of $\pi_0$ meson \cite{Brauner:2016pko,Son:2007ny,Eto:2012qd} (see also refs.~\cite{Kawaguchi:2018fpi,Chen:2021vou,Gronli:2022cri,Yamamoto:2015maz,Brauner:2017mui,Brauner:2021sci,Evans:2022hwr} for related works).
As other examples, the rotation-induced CSL~\cite{Huang:2017pqe,Nishimura:2020odq,Eto:2021gyy,Chen:2021aiq}, Floquet-engineered CSL~\cite{Yamada:2021jhy}, the CSL in QCD-like theory~\cite{Brauner:2019aid,Brauner:2019rjg}, and the CSL-like pattern formation via nonequilibrium process \cite{Yamamoto:2018hdy}
have also been discussed.

There are three important elements in theories that have the CSL state as the  ground state, 
i) a pNGB (referring to $\phi$) associated with the spontaneous breaking of 
a global symmetry $G$,
ii) an explicit breaking of the symmetry $G$ which leads to a periodic cosine-type potential,  
and iii) a total derivative term of $\phi$.
\footnote{In the main text, we suppose that
the origin of the total derivative term 
is the Chiral Separation Effect (CSE)~\cite{Vilenkin:1980fu,Son:2004tq,Metlitski:2005pr,Fukushima:2008xe},
but it does not always have to be the case. 
Depending on the underlying physics of the system, the origin of the total derivative term varies. For example, 
the Chiral Vortical Effect~\cite{Vilenkin:1979ui,Vilenkin:1980zv,Landsteiner:2011cp,Landsteiner:2012kd,Landsteiner:2016led,Son:2009tf}  leads to 
a rotational counterpart of the total derivative term 
discussed in refs.~\cite{Huang:2017pqe,Nishimura:2020odq}. In chiral magnets, the total derivative term of the magnon originates from the so-called Dzyaloshinskii-Moriya (DM) interaction \cite{DZYALOSHINSKY1958241,PhysRev.120.91}.
}
With these ingredients, the low energy effective theory for the pNGB  
becomes the Sine-Gordon theory with the total derivative term.
In the usual Sine-Gordon theory, the domain wall interpolating between the minima of the cosine-type potential appears.
Such a field configuration is topologically protected, but energetically 
unfavorable compared to the trivial vacuum state.
The total derivative term decreases the tension of the domain wall, 
and it can be negative for
a sufficiently large external background field. 
Consequently, the domain wall is energetically more favorable than the vacuum state.

A fact that a field configuration is energetically favorable does not mean that such a configuration is 
formed instantaneously.
Noting that the domain walls are 
topologically stable, we expect that the CSL form with a non-trivial dynamics associated with a change of topological numbers.
The first way  to form the CSL  that one can imagine would be the Kibble-Zurek mechanism~\cite{Kibble:1976sj,Zurek:1985qw}. 
However, it requires the symmetry restoration with {\it e.g.}, a high-temperature environment 
and does not describe the defect formation from the vacuum state.
In order to see the CSL formation at zero-temperature, we need to describe it as a quantum tunneling, 
which we would like to explore in the present paper.
\footnote{See also ref.~\cite{paterson2019order} for the phenomenological study on the CSL formation in the chiral magnets.} 
Generally, it is difficult to calculate the nucleation rate of topological defects starting from the infinite dimensional quantum field theory. 
However, in ref.~\cite{Basu:1991ig} a genius way to calculate it by describing the system 
in terms of  the Nambu-Goto (NG) theory so that the problem becomes one-dimensional quantum mechanics
has been proposed.
Although this method was originally invented for the quantum creation of the topological defects 
in the (quasi) de Sitter background,
it has been noticed that it can be useful for other phenomena such as the Schwinger effect or the vacuum decay through
 the bubble nucleation~\cite{Ai:2020vhx,Hayashi:2021kro}.
In this paper, with adopting this method, we evaluate the single domain wall as well as the CSL formation rate.
Noting that the formation of a domain wall spread to the spatial infinite would be unlikely to occur 
from the viewpoint of
causality, we describe the system as a domain-wall disk surrounded by a vortex or string loop.
The dynamics of the domain-wall disk can be then described by quantum mechanics for the radius of the disk, $R$. 
We show that when the tension of the domain wall is negative, the domain-wall disk with the radius $R_2$ (see eq.~(\ref{R2})) nucleates via quantum tunneling whose rate depends on the external magnetic field strength
as well as the chemical potential of the system.
We find that the nucleation rate is exponentially suppressed below the critical magnetic field amplitude $B_c$ (see eq.~(\ref{Bnucl})).
Therefore, we emphasize that, even when the CSL state is more energetically favorable than the vacuum, the formation of the domain wall is not promoted instantaneously as long as $B<B_{c}$.

As concrete examples, 
we consider the following two realistic physical systems to which the effective theory is expected to be applied, namely, i) the neutral pion in the two-flavor Quantum Chromo Dynamics (QCD) and ii) Axion-Like Particles (ALPs)
~\cite{Svrcek:2006yi,Conlon:2006tq,Arvanitaki:2009fg,Acharya:2010zx,Higaki:2011me,Cicoli:2012sz,Marsh:2015xka} at a finite chemical potential
under an external magnetic field. In the former case,  
the properties of the CSL composed of the 
$\pi_0$ meson associated with the spontaneous chiral symmetry breaking have been studied 
in ref.~\cite{Brauner:2016pko}. 
There it has been pointed out that a too strong magnetic field causes a tachyonic instability 
of the charged pion fluctuation, which determines the upper bound of the magnetic field strength for the CSL. 
We find that the pion domain-wall as well as the pion CSL formation rate becomes larger for stronger magnetic fields, 
but at the upper bound of the magnetic field strength suggested by the stability against the charged pion fluctuation
the formation rate is still exponentially suppressed even though
the CSL state is more energetically favorable than the QCD vacuum.
In the latter case with an ALP, we find for the first time that the total derivative term required for the CSL appears in the effective field 
theory from the Chiral Magnetic Effect (CME)~\cite{Vilenkin:1980fu,Fukushima:2008xe} and the CSL is a ground state
for a sufficiently large external magnetic field. 
Compared to the case of the neutral pion,  we do not have to worry about the instability of associated fields in the ALP sector if they are sufficiently heavy, if any.
As a result, there will be no exponential suppression factor in the nucleation rate if the magnetic field 
and the chemical potential is slightly larger than the axion decay constant.

This paper is organized as follows.
In the next section, we review the CSL and describe it in terms of the NG-type action. 
Then we estimate the formation rate of a single domain-wall as well as the CSL. 
In Sec.~\ref{sec:applications}, we apply the results of Sec.~\ref{sec:nuck_CSL} 
to the physically well-motivated system, that is, the neutral pion in the two-flavor QCD and the ALP. 
Sec.~\ref{sec:Dis_Concl} is devoted for the conclusion and discussion.

\section{Nucleation rate of the chiral soliton lattice} \label{sec:nuck_CSL}
This section considers the nucleation of the CSL.
Since the CSL is a topological defect, it is topologically distinct with the trivial field configuration
and can be generated by the Kibble-Zurek mechanism~\cite{Kibble:1976sj,Zurek:1985qw} or through quantum tunneling. 
Here we are interested in its nucleation at zero temperature, we shall study the latter.
While a quantum field theoretic investigation is quite involved, 
here we adopt a quantum mechanical approach where the CSL is described by the Nambu-Goto-type action 
with a one dimensional parameter, that is, the radius of the wall, as has been first studied in ref.~\cite{Basu:1991ig} 
and also recently adopted in refs.~\cite{Ai:2020vhx,Hayashi:2021kro}. 
In the following, we first review the nature of the CSL and develop its description by the Nambu-Goto-type action. 
Then we evaluate the nucleation rate  for the single disk and  the CSL.

\subsection{Chiral Soliton Lattice (CSL)} \label{sec:CSL_and_NG}
We start with reviewing the CSL 
with a simple toy model. 
We consider a system in the uniform background magnetic field ${\bm B}$ with the spontaneous chiral symmetry breaking 
whose low-energy dynamics can be described by the effective field theory of pNGB $\phi$, 
which will be identified as a neutral pion or an ALP.
Hereafter we neglect fields other than pNGB that may exist in theories of our interest
unless otherwise stated, since they are supposed not to be directly associated with the CSL. For instance, we focus on a situation in which they are much heavier than pNGB. 
The effective Lagrangian 
of the pNGB 
up to the leading order is that of the Sine-Gordon theory, given by
\begin{gather}
   \mathcal{L}=\frac{f^2}{2}(\del_{\mu}\phi)^2 + f^2m^2 ({\rm cos}\phi-1) + \frac{\mu}{4\pi^2}{\bm B}\cdot {\bm \nabla}\phi \label{original lagrangian} \,,
\end{gather}
where $f$ and $m$ are the decay constant and the mass of the pNGB, respectively.
The first term is the kinetic term,
and the second is the 
potential associated with the explicit chiral symmetry breaking. The constant $f^2m^2$ is an offset to zero the energy of the vacuum at $\phi=0$.
The third term is introduced, for example,  in the following reason.
Here we suppose a system with $N_{\textrm{f}}$ species of massless U(1) charged Dirac fermions $\psi_i$ with their  chemical potentials $\mu$ being coupled to the conversed charge $j^0 \equiv \sum_i^{N_f} \bar{\psi_i}\gamma^0\psi_i$.
This system exhibits the so-called chiral separation effect (CSE), 
where  the axial vector currents $\bm{j}_5  {\equiv \sum_i^{N_f} {\bar \psi}_i \gamma^5 {\bm \gamma} \psi_i}$ are induced in the direction of the magnetic field as~\cite{Vilenkin:1980fu,Son:2004tq,Metlitski:2005pr,Fukushima:2008xe} 
\begin{gather}
    \bm{j}_5=\frac{\mu}{2\pi^2}\bm{B}  \,,
\end{gather}
where the gauge coupling is absorbed by the magnetic field. We shall note that
this transport phenomenon is 
related to the chiral anomaly~\cite{Fukushima:2008xe,Son:2009tf,Son:2012wh} and does not receive any renormalization.
Hence, the CSE appears independently of the energy scale. In particular, in phases where the chiral symmetry is spontaneously broken, 
the anomaly matching condition tells that
the CSE should be reproduced by the effective interaction of the resultant
pNGBs, such as the mesons and axions,
which leads to 
the third term in eq.~(\ref{original lagrangian}).
Note that under the chiral transformation $\psi \rightarrow e^{i \gamma^5 \theta} \psi$
the NGB in the low energy effective theory transforms as $\phi \rightarrow \phi+2 \theta$.

Under the chiral rotation with a position-dependent $\theta = \theta(x)$,
the action changes as 
$\delta S = \int d^4x (\partial_\mu \theta (x)) j^\mu_5 = 
\frac{1}{2}\int d^4x (\partial_\mu \delta \phi) j^\mu_5 $
from the Noether's theorem.
Thus, we have $S=\frac{1}{2}\int d^4x (\partial_\mu \phi) j^\mu_5$ in the low energy regime.

The effective Hamiltonian coming from eq.~(\ref{original lagrangian}) is now given as,
\begin{gather}
    \mathcal{H}=\frac{f^2}{2}\left[(\del_t\phi)^2+(\del_x\phi)^2+(\del_y\phi)^2+(\del_z\phi)^2 \right]
    +f^2m^2(1-\textrm{cos}\phi) - \frac{\mu}{4\pi^2}B\del_z\phi \,,
\end{gather}
where we have set $\bm{B}=(0,0,B)$ without loss of generality. 
In order to minimize the Hamiltonian, 
$\phi$ should not depend on $t$, $x$ and $y$ so that it 
becomes
\begin{gather}
    \mathcal{H}=\frac{f^2}{2}(\del_z\phi)^2
    +f^2m^2(1-\textrm{cos}\phi) - \frac{\mu}{4\pi^2}B\del_z\phi \label{hamiltonian_1D} \,.
\end{gather}
Note that in  
the equation of motion, 
\begin{gather}
    \del_z^2\phi=m^2\textrm{sin}\phi 
    \label{sine-gordon-eq} \,, 
\end{gather}
the topological term does not appear. Its
solution with the boundary condition $\phi(-\infty)=0\,, \phi(\infty)=2\pi$
is the so-called domain wall:
\begin{gather}
    \phi_{\mathrm{DW}} (mz)=4\textrm{arctan}[\exp(mz)] \label{sol_wall} \,.
\end{gather}
Substituting eq.~(\ref{sol_wall}) into eq.~(\ref{hamiltonian_1D}),
the energy of this single soliton per unit area in the $x$-$y$ plane  can be calculated as
\begin{gather}
    \mathcal{E}_{\rm DW} 
    \equiv \int_{-\infty}^\infty dz \mathcal{H} [\phi_{\mathrm{DW}} (z)]  = 8mf^2 - \frac{\mu B}{2\pi} \,.
\end{gather}
We can see that while the topological term does not appear in the equation of motion, it appears in the energy.
In the absence of topological term,
the energy of this single domain wall is, of course, larger than the vacuum energy $0$ (or the energy of the trivial configuration $\phi=0$).
On the other hand, as the magnetic field $B$ increases and exceeds the critical magnetic field,
\begin{gather}
    B_{\textrm{DW}}=\frac{16\pi mf^2}{\mu} \label{critical-mag} \,,
\end{gather}
the sign of $\mathcal{E}_1$ changes from positive to negative.
In such a case, the domain wall is not only topologically stable but also energetically favorable, which opens up the possibility to form it as a quantum vacuum decay. 

The fact that the single domain wall is more stable than the vacuum state $\phi=0$
tells that the ground state of the system at $B>B_{\textrm{DW}}$ would be 
a parallel stack of this domain walls.
Since the wall-wall interaction would generate a positive energy, there should be an appropriate distance 
between the walls to minimize the energy. 
The configuration of such a state,  
called the chiral soliton lattice (CSL), is determined as follows. 
One of the general solutions of eq.~(\ref{sine-gordon-eq}) that satisfies $\phi(0)=-\pi$ is given as 
\begin{gather}
    \int_{0}^{\phi/2+\pi/2}d\theta \, \frac{1}{\sqrt{1-k^2\textrm{sin}^2\theta}} =
    \frac{zm}{k} \,.
\end{gather}
Note that both sides of the equation above are zero when $\phi=-\pi$ and $z=0$, and
the solution is a function of $zm/k$, with $k\, (0<k<1)$ being the parameter that characterizes the solution.
In terms of  the Jacobi's amplitude function $\textrm{am}$, it is rewritten as
\begin{gather}
  \frac{1}{2}\left(\phi_k\left(\frac{zm}{k}\right)  + \pi \right)
  = \textrm{am}\left(\frac{zm}{k},k\right) \,. \label{layersol}
\end{gather}
Equivalently, $\phi$ is also given by the Jacobi's elliptic function:
\begin{gather}
\textrm{cos}
\left(\frac{1}{2}\phi_k\left(\frac{zm}{k}\right)\right)= \textrm{sn}\left(\frac{zm}{k},k\right).
\end{gather}
In the equations above, it has been assumed that $\partial_z \phi >0$ and $\mu B >0 $ without a loss of generality, such that $\mu B \partial_z \phi > 0$.
\footnote{
Instead, a solution is given by $\frac{1}{2}\left(\phi_k\left(\frac{zm}{k}\right)  + \pi \right) = - \textrm{am}\left(\frac{zm}{k},k\right)$ when $\partial_z \phi <0$ and $\mu B <0$.
}
Note that the parameter $k$ is now identified as the elliptic modulus.
Using the complete elliptic integral of the first kind $K(k)$, 
the periodicity of this solution is given by 
\begin{gather}
    l=\frac{2kK(k)}{m} \,, \qquad {\rm where} \quad
    2K(k)=   \int_{0}^{\pi}d\theta \, \frac{1}{\sqrt{1-k^2\textrm{sin}^2\theta}} \,,
\end{gather}
with shifts of $\Delta \phi = 2\pi$ and $\Delta z = l$, 
which corresponds to the distance between the walls.
The minimization of 
the energy of each wall with period $l=l(k)$ per unit area in the $xy$ plane,
\begin{equation}
{\cal E}_{\rm CSL}
\equiv   \int_{-l(k)/2}^{l(k)/2}  dz {\cal H}[\phi_k(z)]
= 4mf^2\left[ \frac{2E(k)}{k} + \left(k -\frac{1}{k} \right)K(k) \right] - \frac{\mu B}{2\pi}
\,,
\label{E-density}
\end{equation}
optimizes the free parameter $k$ for given parameters, $\mu, B, m, f$.
The optimal condition is given by~\cite{Brauner:2016pko}
\begin{gather}
  \frac{E(k)}{k}=\frac{\mu B}{16\pi mf^2} \label{minimization_condition} \,,
\end{gather}
where $E(k)$ is the complete elliptic integral of the second kind, 
which determines the elliptic modulus for the CSL, $k=k_\mathrm{CSL}$, and hence the wall distance $l$ for given parameters, $\mu, B, m, f$. Note that
its left-hand side is bounded from below as $E(k)/k> 1\, (0<k<1)$.
Not the vacuum solution but the CSL is the ground state if eq.~\eqref{minimization_condition}
has a solution for $0<k<1$, and 
hence the critical magnetic field strength for the CSL to be the ground state, $B_\mathrm{CSL}$, is equal to eq.~(\ref{critical-mag}), 
\begin{equation}
B_\mathrm{CSL} = B_\mathrm{DW} = \frac{16 \pi m f^2}{\mu}.\label{criticalMFstate}
\end{equation}
\begin{figure}[tb]
    \centering
    \includegraphics[width=7.5cm]{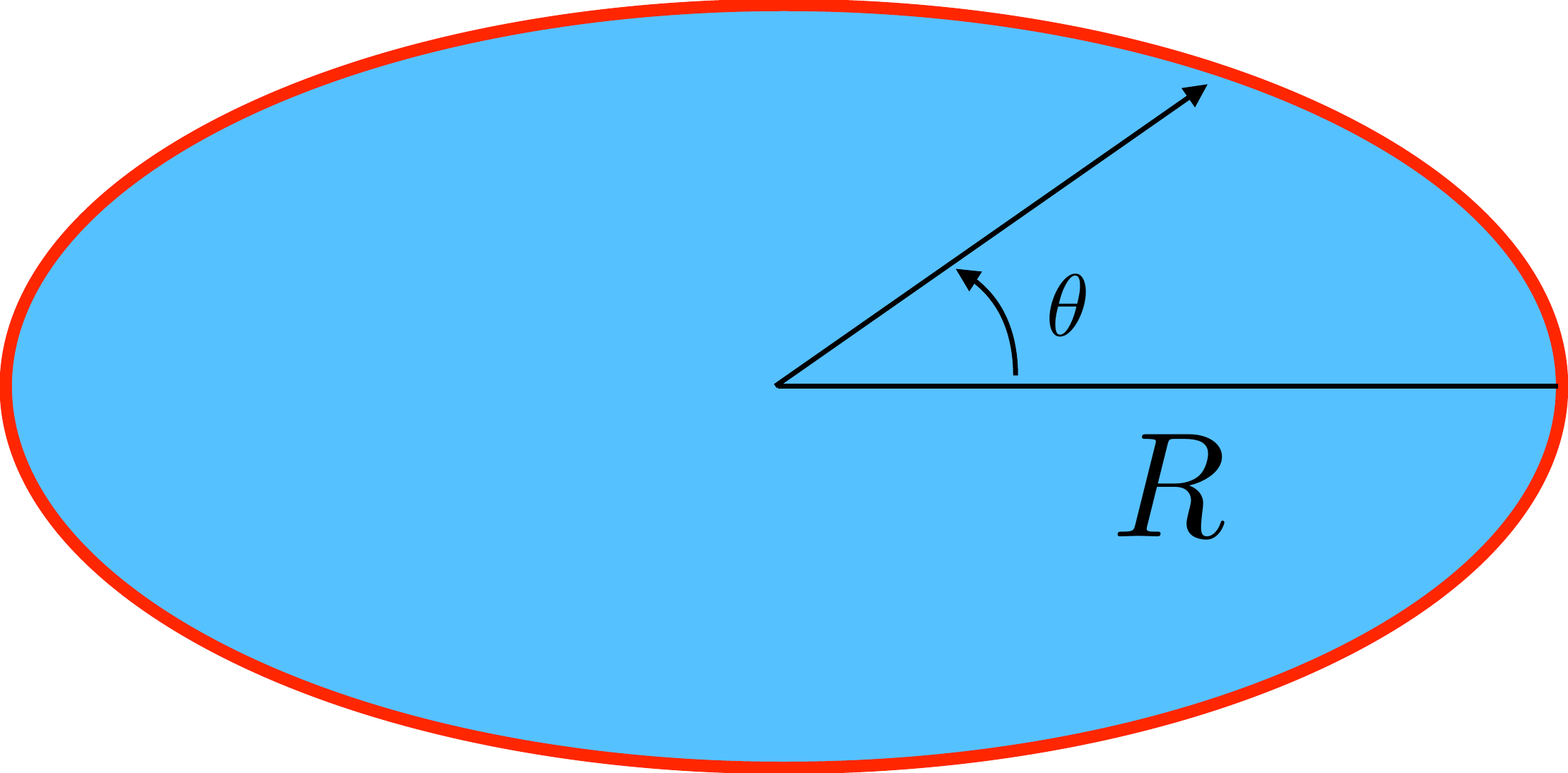}
    \caption{Schematic picture of the domain wall disc with the string on the edge.
  The blue part corresponds to the domain wall, the red the string. The parameter $\theta$ parametrizes the string on the edge of the wall.}
    \label{fig:wall_and_string}
\end{figure}
In a large magnetic field limit of $B \gg B_\mathrm{CSL}~(k \to 0)$, 
in which pNGB mass is neglected, the solution minimizing the Hamiltonian is 
given by 
\begin{equation}
\phi(z) \simeq \frac{\mu B z}{4\pi^2 f^2}.
\end{equation}
Then, the period reads
\begin{equation}
l \simeq \frac{8\pi^3 f^2}{\mu B} .
\end{equation}

\subsection{Nambu-Goto action for the wall-string system}
We have reviewed that the ground state of the Sine-Gordon theory becomes the CSL state due to the topological term. 
This suggests that the CSL with finite topological numbers can be formed from the vacuum state. 
Here, we argue the domain wall formation through
the quantum-mechanical tunneling.
It is easy to imagine that not the infinite domain wall in $x$-$y$ plane is generated instantaneously 
but a disk-like domain wall with a finite size surrounded by a vortex or a string loop~\cite{Fukushima:2018ohd,Kaplan:2001hh,Buckley:2002ur}, as shown in Fig.~\ref{fig:wall_and_string}, is generated first and eventually expands.
Such a system can be formed through spontaneous breaking of the
chiral symmetry in the QCD~\cite{Fukushima:2018ohd,Kaplan:2001hh,Buckley:2002ur}
or a that of $U(1)_{\rm PQ}$ in models with ALPs 
\cite{Zeldovich:1974uw,Kibble:1982dd,Vilenkin:1982ks}.
Infinite dimensional quantum field theoretic approach on the formation of such a field configuration is quite involved, 
and hence we adopt the thin-wall approximation so that the field configuration can be described by one-dimensional Nambu-Goto-like action
as has been studied in refs.~\cite{Basu:1991ig,Ai:2020vhx,Hayashi:2021kro}.
In this subsection, we construct this one-dimensional effective action for the wall-string system.

Here we consider the wall-string system in the Minkowski background, 
\begin{equation}
\rmd s^2 = \eta_{\mu\nu} \rmd x^\mu \rmd x^\nu = -\rmd t^2+ \rmd \Omega_3, 
\end{equation}
where $\rmd\Omega_3$ is the line element on the three-dimensional Euclidean space. 
The Nambu-Goto like effective action~\cite{Vilenkin:2000jqa} of the domain-wall disk surrounded by an edge
can be divided into the wall (void) part and string (edge) part as
\begin{equation}
S = S_\mathrm{wall} + S_\mathrm{string} = -\sigma \int_\mathcal{W} d^3 \zeta \sqrt{\mathrm{det} h_{ab}} - T \int _{\partial \mathcal{W}} d^2 \xi \sqrt{- \mathrm{det} \gamma_{\alpha \beta}}, 
\end{equation}
where $\sigma$ and $T$ are the wall and string tensions, $h_{ab}$ and $\gamma_{\alpha \beta}$ are 
the induced metrics on the three-dimensional world volume for the wall and two-dimensional world sheet for the string, respectively.

Let us first investigate the former, the wall action.
The induced metric on the wall is given by
\begin{gather}
h_{ab} = \eta_{\mu \nu} \frac{\del X^{\mu}}{\del \zeta^a} \frac{\del X^{\nu}}{\del \zeta^b} \,, 
\end{gather}
where $X^\mu(\zeta)$ is the three-dimensional world volume of the domain wall embedded in the bulk spacetime, ${\mathcal W}$,
with $\zeta^a \ (a=0,1,2)$ being the coordinate covering ${\mathcal W}$.
Here $\zeta^0$ is chosen to be timelike whereas $\zeta^{1,2}$ are chosen to be spacelike.
Since we are interested in the disk-like configuration where the wall is perpendicular to the $z$ axis while
rotational symmetric in the $x$-$y$ plane, it is natural to take the coordinate on the wall to be
$\{\zeta^a\} = \{\tau, \chi, \theta\}$ and the trajectory to be $\{X^\mu \} = \{t(\zeta), r(\zeta), \Theta(\zeta), z=z_0\}$. 
Fixing the gauge as $\tau = t$, $\chi=r$, and $\theta=\Theta$, the induced metric is given as
\begin{equation}
h_{ab} = \mathrm{diag} ( 1, -1, -r^2), 
\end{equation}
where we have taken the cylindrical coordinate system for the background Minkowski metric,
$ds^2 = dt^2 - dr^2 -r^2 d\theta^2 -dz^2$. 
The tension of the wall is calculated by integrating the energy density along the $z$-axis. 
For the single domain wall solution eq.~\eqref{sol_wall}, we obtain
\begin{equation}
\sigma \equiv \int dz {\cal H}=  8 m f^2 - \frac{\mu B}{2\pi}. 
\end{equation}
Note that the tension is negative at $B>B_\mathrm{DW}=B_{\textrm{CSL}}$.
Since we consider the case where the disk radius can change with time, the domain of $r$ is given as $0<r<R(\tau)$. 
Consequently, we get the Nambu-Goto-like action for the domain-wall disk as
\begin{equation}
S_\mathrm{wall}  = -\sigma \int_\mathcal{W} d^3 \zeta \sqrt{\mathrm{det} h_{ab}}  = - \sigma \int d\tau \int_0^{R(\tau)} dr \int_ 0^{2 \pi} d \theta r = -  \pi \sigma \int d \tau R(\tau)^2. 
\end{equation}

Similarly to the wall part, 
the induced metric on the string worldsheet is given by
 \begin{gather}
 {\gamma}_{\alpha \beta} = \eta_{\mu \nu} \frac{\del X^{\mu}}{\del \xi^{\alpha}} \frac{\del X^{\nu}}{\del \xi^{\beta}} \, ,
\end{gather}
where $X^\mu = X^\mu(\xi)$ describes the two-dimensional string worldsheet 
$\partial {\cal W}$ with the radius $R$ embedded in the bulk spacetime as shown in Fig.~\ref{fig:wall_and_string} so that we can take $\{X^\mu \} = \{t(\xi), R(t(\xi)),\Theta(\xi), z=z_0\}$, with 
$\xi^{\alpha}\, (\alpha=0,1)$ being the coordinate covering the $\partial {\cal W}$, 
$\{\xi^a\} = \{\tau, \theta\}$. 
We gauge-fix the coordinates as $\tau =t $ and $\theta = \Theta$, and emphasize that $R$ is a dynamical and depends on $t$: $R=R(t)=R(\tau)$ on the $\partial {\cal W}$.
Then, its explicit form 
is calculated as
\begin{gather}
\gamma_{\alpha \beta} = {\rm diag}(1-\dot{R}^2, -R^2) ,
\end{gather}
or
\begin{gather}
 \rmd s^2_{\rm worldsheet} 
= (1-\dot{R}^2) \rmd \tau^2 - R^2 \rmd \theta^2 ,
\end{gather}
where the dot represents the derivative with respect to $\tau$.
The string action is proportional to the worldsheet area:
\begin{gather}
S_{\rm string} = -T \int \rmd \tau \rmd \theta 
\sqrt{-{\gamma}}
= -2\pi T \int \rmd \tau R\sqrt{1-\dot{R}^2} \,,
\end{gather}
where the string tension evaluated at far from the string core, $T$, is 
the summation of the potential energy and the gradient energy of the radial direction of the
symmetry breaking field or the order parameter and is
calculated as~\cite{Hindmarsh:1994re,Chatterjee:2019rch} 
\begin{gather}
    T \sim 2\pi \times 2f^2 \times \ln \frac{R_{\textrm{c}}}{r_{\textrm{c}}} \,,
\end{gather}
where $R_{\textrm{c}}$ and $r_{\textrm{c}}\sim f^{-1}$ are cutoff length and string core size, respectively.
Here we have only taken into account the classical contribution from the radial direction of the 
symmetry breaking field that does not couple to the magnetic field. Thus
we regard that the string tension is independent of the magnetic field throughout this paper. 
Once we take into account light matter fields charged under the U(1) gauge theory, however, there can arise a magnetic-field-dependent 
contribution to the string tension through, e.g., the fermion zero modes, 
but it is UV model-dependent.
Since the classical part would generally dominate over the whole contribution in the tension, we expect that our analysis would be valid, even if 
the string tension is affected by the magnetic field, depending on the detail of the model.
The potential energy stored inside of the core  is estimated as of order
$f^2$, whereas the gradient energy gives the logarithmic factor.
A naive choice of $R_\mathrm{c}$ would be the disk radius $R_\mathrm{d}$. 
However, in the wall-string system the field configuration around the string is non-trivial and different from the string 
without being attached to a wall, 
and hence careful numerical study is needed for the estimate for the cutoff length or the gradient energy. Instead, here we take
$R_\mathrm{c}$ 
as a parameter independent of the disk radius $R_\mathrm{d}$
throughout this paper, leaving its precise determination for future study.
In ref.~\cite{Eto:2022lhu}, the authors have numerically shown
$T \sim f^2 \sim {\rm const.} $, independent of the disk radius as well as the magnetic field strength, at a distance from a domain wall disk bounded by a string loop.
That is because unlike the strings without being attached to a wall, the field configuration appears to exponentially come close to that of the vacuum, in which there exists no solitons, as being away from such a disk, regardless of the value of the wall radius. 
In this case, one length scale that could be related to the string configuration is the thickness of the wall,  $r_\mathrm{w} \sim 1/m$.
Hence another plausible estimate for $R_\mathrm{c}$ would be the thickness of the wall, $r_\mathrm{w}$, which is independent of $R_\mathrm{d}$, but careful investigations are needed to confirm this ansatz.
\footnote{
We thank M. Eto and M. Nitta for pointing out this fact.
}

\subsection{Calculation of the decay rate}\label{sec:cal_decay_rate}
Now we are ready to evaluate the quantum mechanical disk nucleation rate
with the help of the Nambu-Goto-like effective action.
As we have discussed the field configurations for the single domain-wall disk and the CSL-like disk layer are different, we need carefully to take care of the difference. 
We will first examine the single disk nucleation and then discuss the CSL formation.

\subsubsection{Nucleation of a single domain wall disk from the vacuum} \label{sec:creation1}
\begin{figure}[tp]
    \centering
    \includegraphics[width=15.0cm]{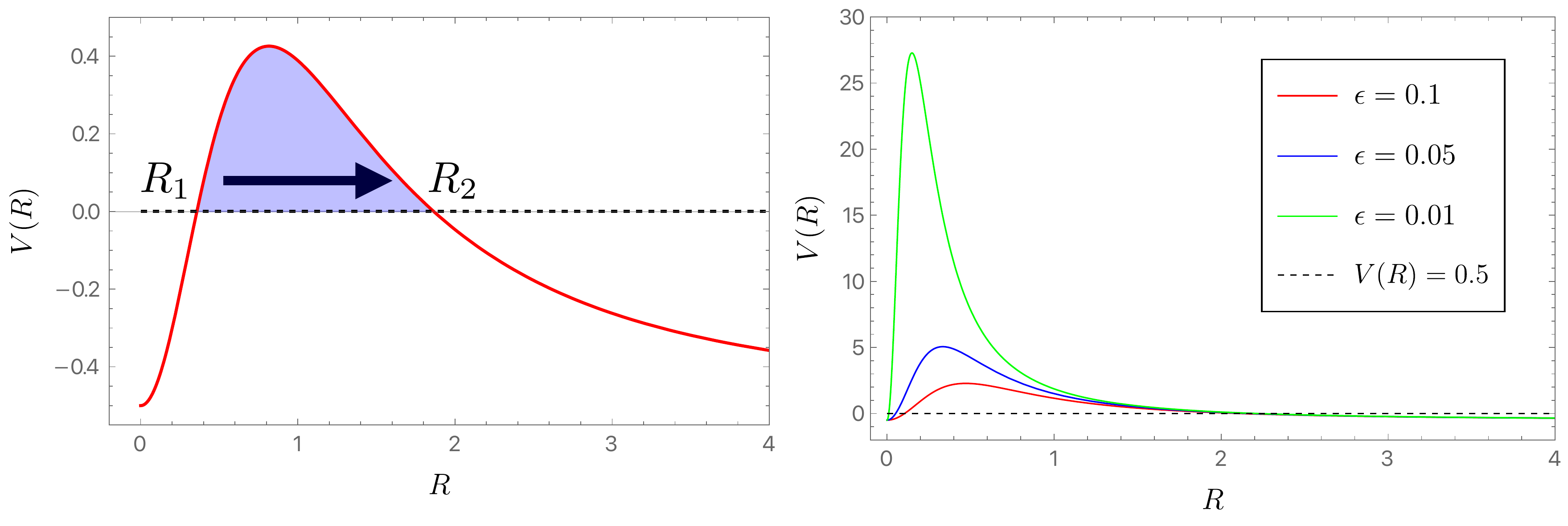}
    \caption{
    (\textit{Left panel}) Typical shape of the potential $V(R)$
    as a function of $R$  is shown in the solid red line ($\sigma/T=-0.9$ and $\epsilon=0.3$). 
    For $\sigma<0$ the barrier height is finite.
    The black dotted one represents the line for $V(R=0)$.
    $V(R)$ is positive in the blue region at $R_1<R<R_2$, in which region 
    classical dynamics is forbidden.
    (\textit{Right panel}) 
    $\epsilon$-dependence of the shape of the potential $V(R)$ for 
    $\sigma/T=-0.9$ is shown. The potential barrier tends to be higher for smaller $\epsilon$, while the width of barrier $\Delta R \equiv R_2-R_1 \sim R_2 \sim -2T/\sigma$ 
    is insensitive to $\epsilon$. 
    }
    \label{fig:barrier}
\end{figure}

First we consider the tunneling process where a single wall is created from the vacuum.
From the discussion in the previous section, 
the dynamics of a single domain-wall disk is described by one-dimensional 
quantum mechanics of the disk radius $R(\tau)$ with the effective action,
\begin{gather}
S_{\rm tot} = S_{\rm wall} + S_{\rm string} =-\pi \int d \tau \left(2 T R \sqrt{1-{\dot R}^2}\ + \sigma R^2\right) \,.
\end{gather}
In order to study the dynamics of this domain wall system,
it is convenient to use the conserved energy,
\begin{gather}
E =  p {\dot R} 
- L \,,
\end{gather}
where the Lagrangian $L$ can be read from the action $S_{\textrm{tot}}$, $L = -\pi (2TR\sqrt{1-{\dot R}^2}+\sigma R^2)$,  and $p$ is the momentum conjugate to $R$, 
\begin{gather}
p = \frac{\del L}{\del {\dot R}} = \frac{2\pi T R\dot{R}}{\sqrt{1-\dot{R}^2}} \,.
\end{gather}
The conservation law can be rewritten as the following form,
\begin{gather}
{\dot R}^2 + \left[
\frac{R^2}{\left(\epsilon - \frac{\sigma R^2}{2T} \right)^2} - 1
\right] \equiv {\dot R}^2 + 2 V(R) = 0 \,,
\end{gather}
where we have introduced $\epsilon = E/(2\pi T)$.
The first and second terms can be identified as the kinetic and potential terms of $R(t)$, respectively.
Note that $V(R)$ diverges for $\sigma >0$ at $R=R_\mathrm{div} \equiv \sqrt{2 \epsilon T/\sigma}$ with $R_1<R_\mathrm{div}<R_2$ and hence nucleation rate is  expected to be highly suppressed.
We can see that  for $T>-2 \epsilon \sigma$ or $T^2> - \sigma E/\pi$ this potential has a potential barrier between $R=0$ and $\infty$
as shown in Fig.~\ref{fig:barrier}.
Note that for $\sigma<0$, the barrier height is finite and hence
it is possible for the quantum tunneling to happen. 
By solving $V(R)>0$ we find the starting and end points, $R_1$ and $R_2$ as  
\begin{gather}
R_1 = 
\begin{cases}
\frac{T}{\sigma}\left[-1 + \sqrt{1+2\epsilon \sigma/T} \right] & (\epsilon > 0) \\
\frac{T}{\sigma}\left[1 - \sqrt{1+2\epsilon \sigma/T} \right] & (\epsilon < 0)
\end{cases} \,, \qquad
R_2 = \frac{T}{\sigma}\left[-1 - \sqrt{1+2\epsilon \sigma/T} \right] \,,
\end{gather}
where we have taken $\sigma<0$, which corresponds to take $B>B_\mathrm{DW}$.
The structure of the potential tells that 
the disk radius $R$ can change classically at
$0<R<R_1$ and $R>R_2$. 
If the initial disk radius is at $0<R<R_1$, it can expand up to
$R=R_1$ and then recollapses to $R=0$.
On the other hand, 
for the initial condition $R>R_2$,
the disk radius expands towards infinity.

Although the classical dynamics is forbidden in the range $R_1<R<R_2$, 
the domain wall disk can nucleate by the quantum tunneling 
through the potential barrier.
The tunneling probability from $R=R_1$ to $R=R_2$
can be evaluated as follows~\cite{Coleman:1977py}.
By performing a Wick rotation, $\tau \rightarrow -i \tau_\mathrm{E}$, we obtain the Euclidean action, 
\begin{equation}
S_\mathrm{E}[R] = \pi \int d \tau_\mathrm{E}  \left( 2 T R \sqrt{1+\left(\frac{dR}{d \tau_\mathrm{E}}\right)^2}+\sigma R^2 \right). \label{R2}
\end{equation}
The bounce solution for the Euclidean action satisfies the conservation law with the flipped potential as
\begin{equation}
\left(\frac{d R}{d \tau_\mathrm{E}}\right)^2  - 2 V(R)=0, \label{bounce}
\end{equation}
with the boundary condition $V(R_1) = V(R_2) =0$. 
With this solution, we can evaluate the bounce action. 
Since we are interested in the domain-wall disk nucleation from the nothing, 
we take the limit $\epsilon \rightarrow +0$ so that
\begin{gather}
    R_1=0 \,, \qquad
    R_2=-\frac{2T}{\sigma} \,.
\end{gather}
As for the sign of a small $\epsilon$, note that the conserved energy is approximately given by the rest energy of a contribution of the string tension plus the wall tension, $E\simeq 2\pi R T + \pi R^2 \sigma = |\sigma| \pi R ( R_2 - R)$,
when the kinetic energy (expansion of the radius) is small. 
As long as we focus on the tunneling process between $R=0$ and $R = R_2$,
$E \propto \epsilon $ is necessarily positive 
because the positive string tension dominates over the conserved energy 
in the range of such a small $R$. (A large wall with $R >R_2$ cancels also the kinetic energy such that $E \propto \epsilon = 0$ owing to the negative tension.)
See Fig.~\ref{fig:barrier} to show this tendency.
Note that the zero initial radius of the domain-wall disk means that there was no domain wall at all.
By solving the bounce equation for $\epsilon \rightarrow 0$ 
and changing the variables from the Euclidean time $\tau_\mathrm{E}$ to the disk radius $R$,
the bounce action is evaluated as
\begin{equation}
\mathcal{B} = 2 \times 2 \pi T \int_{0}^{R_2} dR\sqrt{R^2-(\sigma R^2/2T)^2} = \frac{16 \pi T^3}{3 \sigma^2}, \label{analytical_form_bounce_action} 
\end{equation}
where the factor 2 comes from the bounce trajectory, $R: 0 \rightarrow R_2 \rightarrow 0$. 
We can see 
that the tunneling action $\mathcal{B}$ remains finite in the limit of
$\epsilon \to 0$.
Consequently, 
the nucleation rate $P$ that
the domain-wall disk with the radius $R=R_2=-2 T/\sigma$ nucleates is given by 
\begin{equation}
P \simeq {\cal A} e^{-{\cal B}} = {\cal A} e^{-16 \pi T^3/3 \sigma^2}, \label{nucleation rate}
\end{equation}
where the prefactor ${\cal A}$ counts for the effects of quantum fluctuation around 
the bounce solution \cite{Coleman:1977py,Callan:1977pt,Basu:1991ig}.

While the domain-wall disk nucleation rate 
is exponentially suppressed for $\mathcal{B}\gg 1$, 
it is unsuppressed for $\mathcal{B}\lesssim 1$.
Noting that $T\simeq 4 \pi f^2 \ln (R_\mathrm{c}/r_\mathrm{c})$ and $\sigma \simeq 8 m f^2 - \mu B/2\pi$, 
we find that for a sufficiently large magnetic field,
\begin{equation}
B>B_c = \frac{16\pi mf^2}{\mu} \left(\frac{(\ln (R_\mathrm{c}/r_\mathrm{c}))^{3/2}}{\sqrt{3}} \frac{4\pi^2f}{m} +1 \right)  (>B_\mathrm{DW})\label{Bnucl} \,,
\end{equation}
the nucleation rate is unsuppressed, which is the main result of the present paper.

The prefactor $\mathcal{A}$ can, in principle, be calculated by evaluating the 
quantum fluctuations around the saddle point solution, but the calculation is rather involved 
in practice. Instead here
we resort to the dimensional analysis.
The action of the domain-wall disk bounded by the string has the time transitional symmetry, and hence the wall production can happen anytime.
The zero mode corresponding to the time-translation symmetry gives a contribution ${\mathcal T}\sqrt{\mathcal{B}/(2\pi)}$ to $\mathcal{A}$.
As with time translational symmetry, the action also has the spatial translational symmetry so that the wall disk production can occur anywhere.  Thus there appears a contribution of $V(\sqrt{\mathcal{B}/(2\pi)})^3$ 
in $\mathcal{A}$.
Taking into account the 
characteristic length scale of the domain-wall disk, $R_{\rm d}= R_2= |2T/\sigma|$, 
we estimate ${\cal A}$ as 
\begin{gather}
\mathcal{A} \propto \left(\frac{\mathcal{B}}{2\pi} \right)^2\frac{1}{R_{\rm d}^4} \,.
\end{gather}

Before proceeding, let us comment on the assumption we have made.
In the above discussion, we have assumed that the system is well described 
by the domain-wall disk surrounded by the string.
In other words, we have assumed that the thickness of the wall $r_\mathrm{w}$ and the radius of the string $r_\mathrm{c}$
are infinitely thin.
The former is evaluated as $r_\mathrm{w} \sim 1/m$ while the latter is evaluated as $r_\mathrm{c} \sim 1/f$. 
The thin-wall/string approximation is valid if they are smaller than the disk radius $R_\mathrm{d}$. Noting that $R_\mathrm{d} = 2 T/|\sigma|$, the conditions read
\begin{align}
    \frac{r_{\textrm{c}}}{R_{\mathrm{d}}}
    &\sim \frac{1/f}{2 \times4\pi f^2 \ln (R_\mathrm{c}/r_\mathrm{c})/
 (8mf^2|1-B/B_\mathrm{DW}|)}
    \sim \frac{m}{\pi f \ln (R_\mathrm{c}/r_\mathrm{c})}\left(
    \frac{B}{B_\mathrm{DW}}
    -1 \right) \ll 1 \label{validity_of_NGaction1} \,, \\ 
       \frac{r_{\textrm{w}}}{R_{\mathrm{d}}} 
    &\sim \frac{1/m}{2 \times 4\pi f^2 \ln (R_\mathrm{c}/r_\mathrm{c})/
 (8mf^2|1-B/B_\mathrm{DW}|)}
    \sim \frac{1}{\pi  \ln (R_\mathrm{c}/r_\mathrm{c})}\left(
    \frac{B}{B_\mathrm{DW}}
    -1 \right) \ll 1 \label{validity_of_NGaction2} \,.
\end{align}
With estimating the logarithmic factor with the cutoff length for the string 
to be of order of the unity,
we find that the thin-wall approximation is valid for the magnetic field strength slightly 
larger than $B_\mathrm{DW}$.
Strictly speaking, the wall disk nucleation rate evaluated in the above 
cannot be used for much larger magnetic fields. 
In particular, the critical magnetic field for the bubble nucleation $B_c$ (eq.~\eqref{Bnucl})
is likely much larger than $B_\mathrm{DW}$ (for example, $B_c \sim (f/m) B_\mathrm{DW} \gg B_\mathrm{DW}$ for $f \gg m$).
However, in ref.~\cite{Hayashi:2021kro} where the three-dimensional bubble domain wall
formation is studied with a Lorentzian formalism, the exponential suppression factor of the 
nucleation rate for a bubble with a larger radius is turned out to be the same to that 
for the critical bubble. This is understood that the classical expansion
after the quantum tunneling to the critical bubble is evaluated in a completely quantum 
mechanical way. 
We expect the same argument follows so that our estimate of the 
bounce action for the critical disk (with $R_\mathrm{d}=R_\mathrm{c}$)
gives that for the disk with larger radius, where the thin-wall approximation is better. 
Moreover, we do not expect that the nucleation rate for the disk wall with the radius smaller
than the wall thickness is much more enhanced than the estimate in the above, 
since the action of the field configuration of such thick wall disk would be the same order 
to the thin-wall approximation.
Therefore we expect that the bounce action eq.~\eqref{analytical_form_bounce_action}  gives a good estimate for 
the suppression factor for the total wall disk nucleation rate even for 
the thin-wall approximation is not good 
for $R_\mathrm{d}=R_\mathrm{c}$.

\subsubsection{Simultaneously generated domain walls and effects of background domain walls}

\begin{figure}[htbp]
  \begin{center}   \includegraphics[width=8.0cm]{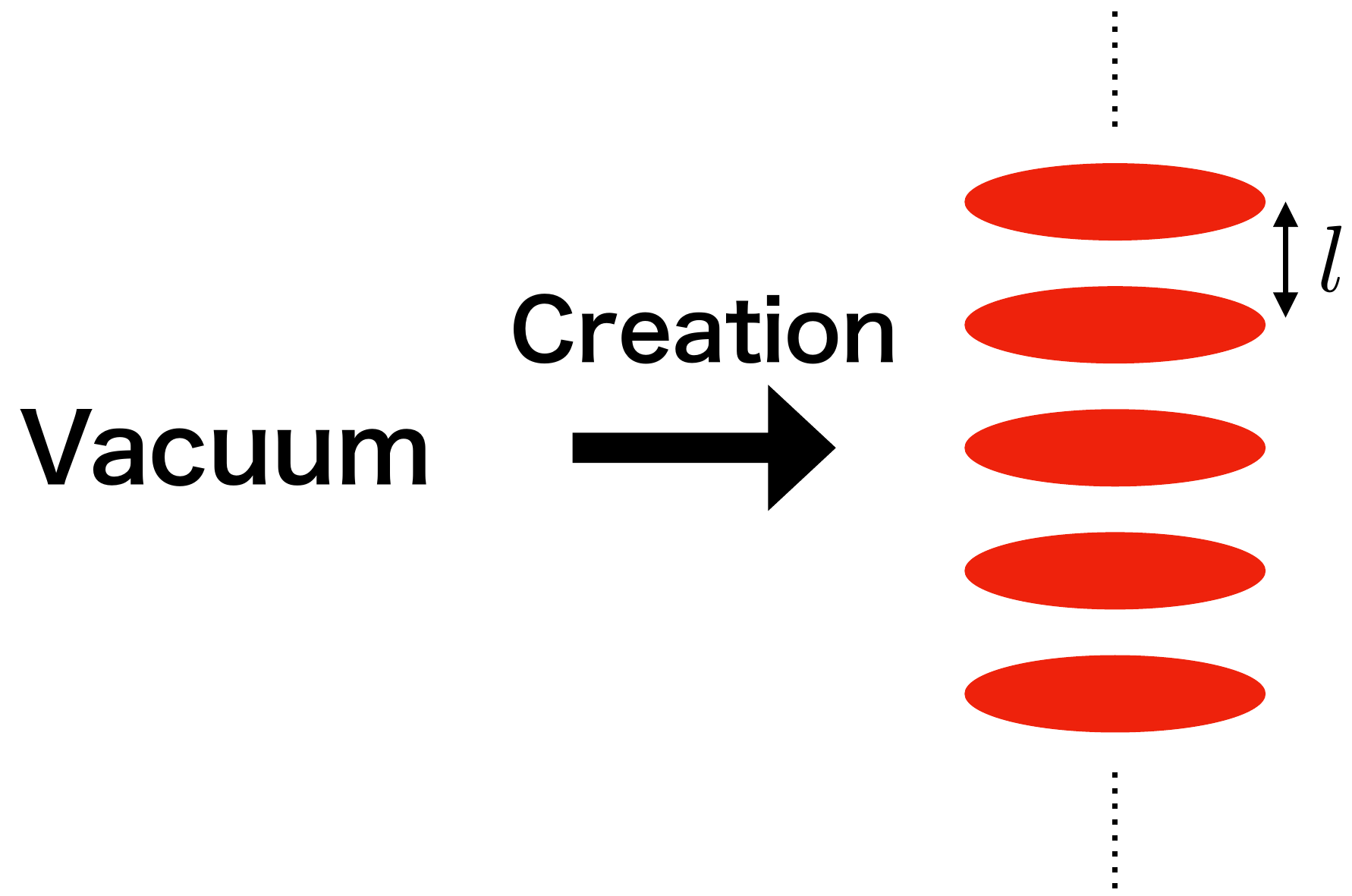}
  \end{center}
  \caption{
  Schematic picture for the simultaneous nucleation of the domain walls  at the interval $l$ is shown.
  We note that $l$ is the lattice space that minimizes the energy density of the domain wall \cite{Brauner:2016pko}.
  }
  \label{fig:walls}
\end{figure}

\begin{figure}[htbp]
    \centering
    \includegraphics[width=11.0cm]{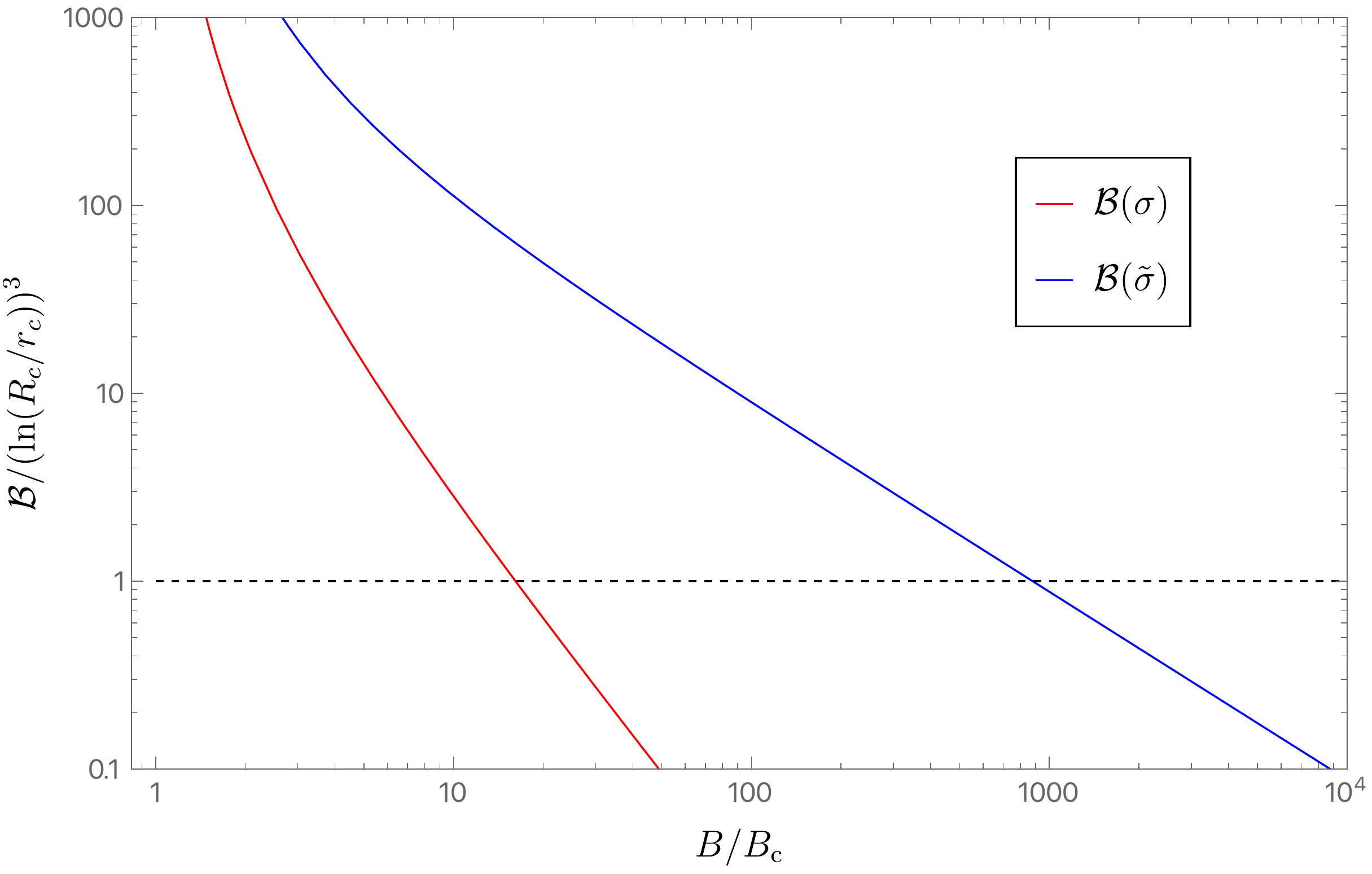}
    \caption{
    The $B$-dependence of the bounce action $\mathcal{B}$ for 
    the single disk (red line) and the CSL (blue line) is shown.
    Since $|\tilde{\sigma}|$ is smaller than $|\sigma|$, 
    the bounce action for the single disk is smaller than that for the CSL.
    The black dotted line shows $\mathcal{B}/(\ln (R_\mathrm{c}/r_\mathrm{c}))^3=1$.
    We emphasize that the point where it interacts with the solid lines gives a critical magnetic field where the nucleation rate is no longer exponentially suppressed.
    }
    \label{fig:bounce_action}
\end{figure}

Next we examine how the CSL forms.
When multiple domain walls are simultaneously created from the vacuum and the CSL system is formed (see Fig.~\ref{fig:walls}), 
a wall feels the existence of the neighbouring walls, and hence
a nucleation rate of one of the walls would be affected. 
In order to take this effect into account, let us
focus on a single wall which constitutes the CSL and consider its tunneling process. 
The critical difference to the single disk formation from the vacuum studied in Sec.~\ref{sec:creation1} is the wall tension, 
because the wall feels the repulsive force originated from the other background walls.
Let us remind that the wall tension for the domain wall layer (eq.~\eqref{layersol})
is given by eq.(\ref{E-density}), 
\begin{gather}
\tilde{\sigma} = 
{\cal E}_{\rm CSL}
= 4mf^2\left[ \frac{2E(k)}{k} + \left(k -\frac{1}{k} \right)K(k) \right] - \frac{\mu B}{2\pi} \,,
\end{gather}
and
the minimization condition that optimizes the parameter $k =k_\mathrm{CSL}$
is given by (\ref{minimization_condition}).
We note that 
$\sigma < \tilde{\sigma} < 0$ 
is satisfied. \footnote{
The derivative of $\sigma-\tilde{\sigma}$ with respect to $\mu B$ is $-E(k)/(4\pi K(k))<0$
and $\lim_{B\to B_{\textrm{c}}}\sigma-\tilde{\sigma} = 0$. Then, we get $\sigma < \tilde{\sigma} $. 
Even though the amplitude of ${\tilde \sigma}$ is smaller than $\sigma$, the number density of 
the wall in $z$-direction is larger, the total energy density in the whole space is smaller for the CSL state.
}This is because repulsive forces working between the walls increases the tension  
and hence an isolated wall is energetically more favorable.
By replacing the wall tension $\sigma$ by ${\tilde \sigma}$ in eq.~\eqref{nucleation rate},
we obtain the wall disk nucleation rate, which
is suppressed by the exponential factor $\rme^{-\mathcal{B}(\tilde \sigma)}$, 
where ${\mathcal B} (\tilde \sigma)= 16\pi T^3/3 \tilde{\sigma}^2$.
Strictly speaking, this rate is those for a single disk nucleated 
at the deficit of the CSL with a distance of $2 l (k_\mathrm{CSL})$, 
we suppose it gives a good approximation for the CSL (with a finite boundary being surrounded by strings) 
formation rate itself.
The $B$-dependence of $\mathcal{B}$ is shown in Fig.~\ref{fig:bounce_action}.
Note that the production of the domain wall is promoted when $\mathcal{B}\lessapprox 1$ is satisfied.
Supposing that $\ln(R_\mathrm{c}/r_\mathrm{c})$ is order of the unity, 
we find that a single wall disk is easily nucleated for $B\gtrsim (10-20) B_\mathrm{DW}$ 
while a fast  CSL formation requires relatively larger magnetic fields, $B\gtrsim 10^3 B_\mathrm{DW}$. 
This is because the amplitude of the wall tension is smaller for the CSL than that of the single disk, 
which is energetically less favored.
One may think that the nucleation rate would not be so suppressed 
compared to the one from the vacuum until the domain wall separation
becomes comparable to the one for the CSL state. 
However, the nucleation rate strongly depends on the (absolute value of) domain wall tension. 
Since the interaction between the domain wall is always repulsive due to the gradient energy of 
the $\phi$ field, and hence the domain wall tension itself increases as we decrease the lattice separation $\ell$ with staying negative for  $B>B_{\textrm{DW}}$. 
Note that even though the CSL state has larger tension, it is energetically favored, 
since the number density of the walls in $z$-direction is larger.
As we have discussed in the previous subsection, for such strong magnetic fields, 
the thin-wall approximation does not hold for the disk with the critical radius 
$R=R_2 = 2 T/|\tilde{\sigma}|$.
However, the CSL with the infinite radius are formed through the expansion of the one with 
a finite radius, for which the thin-wall approximation holds at some point. 
Since the suppression factor of the nucleation rate of the wall disk with larger radius 
is expected to be the same, as has been studied in ref.~\cite{Hayashi:2021kro}, 
the estimate read from Fig.~\ref{fig:bounce_action} would be appropriate to evaluate
the disk/CSL formation rate.

Before concluding this section, let us discuss how the system evolves as a whole.
Once a single domain-wall disk forms quantum mechanically 
with a rate $\propto e^{-{\cal B}(\sigma)}$, 
it expands classically because the potential $V(R)$ monotonically decreases at $R>R_2$.
If another disk has been created in the same $x$-$y$ plane, the disks will eventually collide
and merge each other.
If $\mathcal{B}(\sigma)<1$, the single wall production rate in a $x$-$y$ plane 
is large and the radius of the domain wall becomes effectively infinite in the $x$-$y$ plane quickly.
If it is also the case with $\mathcal{B}({\tilde \sigma})<1$, the domain wall formation
in the $z$ direction is not suppressed in spite of 
the repulsive force of the background walls~\cite{Weinberg:2012pjx}
so that the CSL forms quickly with infinite in the 
$x$-$y$ direction in a similar way to the ordinary liquid/gas transition (in 2-dimension).
On the other hand, if  $\mathcal{B}(\sigma)<1<\mathcal{B}({\tilde \sigma})$, 
the completion of the CSL formation takes time while infinite domain walls with a relatively large distances are formed relatively quickly through the merger of the disks in the same $x$-$y$ plane. Even in the presence of disks, the nucleation rate in the same $x$-$y$ plane is not suppressed compared
to the one from the vacuum and 
they  will merge relatively quickly to form the domain wall with an infinite radius. On the other hand,
due to the repulsive force between walls in the $z$-direction, the nucleated domain walls are hard to merge into other walls along $z$-direction.
Therefore, the bubble collision along $z$-direction is expected to be rare and hence the phase transition from the vacuum to the CSL is harder than that from the vacuum to a wall which is infinitely distant to other walls both in $z$- and $x$-$y$ direction.

\section{Implication for physical specific systems}\label{sec:applications}

The discussions in the previous section is based on a simple toy model only with a pNGB, which catches up the 
basic physics. 
Once we consider more realistic models, some of the details are different, such as the definition of $\mu$ and ${\bm j}_5$, 
while quantitative arguments are possible.
In this section, we take the neutral pion in QCD and axions in the early Universe as examples 
and discuss their phenomenology.

\subsection{QCD} \label{sec:application_qcd}
\begin{figure}[htbp]
    \centering
    \includegraphics[width=11.0cm]{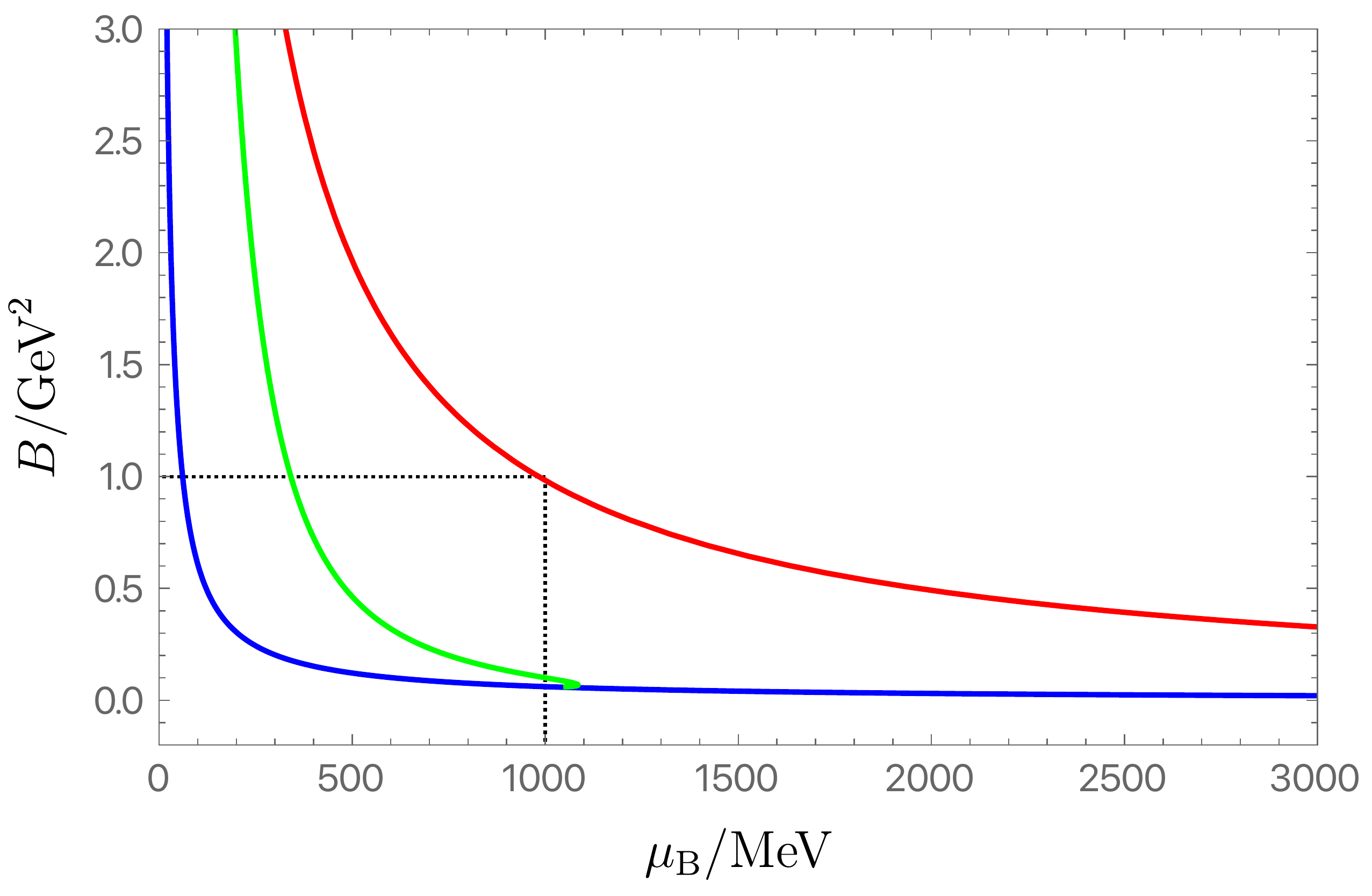}
    \caption{
   The critical magnetic fields for the state, $B^{\pi_0}_{\textrm{CSL}}$ (Blue),  for the BEC, $B_{\textrm{BEC}}^{\pi_{\pm}}$ (Green), and for the single disk nucleation, 
    $B_{\textrm{c}}^{\pi_0}$ (Red) are shown as functions of $\mu_{B}$.
    The region below the blue line indicates the parameter space where the ground state is the QCD vacuum.
    The  area between the blue and green line is the parameter space where the $\pi_0$ CSL state is the ground state, while in the region above the green line the CSL state
    is unstable     and the ground state becomes $\pi_{\pm}$ BEC state.
    In the region above the red line, 
    the nucleation of the domain walls is promoted promptly without the exponential suppression.
    We emphasize, however, this region lies above the green line, where 
    the CSL is no longer the vacuum state, as well as the black dotted line ($B = 1 \mathrm{GeV}^2$ and $\mu_B=1 \mathrm{GeV}$), which indicates the boundary of the region where  
    the chiral perturbation theory is valid. 
    }
    \label{fig:critical_mag_qcd}
\end{figure}
Let us first
apply our result to that in 2-flavor QCD at the finite baryon chemical potential $\mu_{\textrm{B}}$ under the external magnetic field, 
where the low energy effective theory is described by the neutral and charged pions. 
In this case, the relevant effective action for the CSL is described by eq.~\eqref{original lagrangian}
just by replacing $\mu$ with $\mu_B$, $\phi$ with the neutral pion $\pi$, $f$ with 
the pion decay constant $f_\pi$,  and $m$ with the pion mass $m_\pi$.
From the same discussion
in section \ref{sec:CSL_and_NG},
the CSL of the neutral pion becomes the ground state for the magnetic field
larger than
the critical magnetic field, $B_{\textrm{CSL}}^{\pi_0}$ (eq.~\eqref{criticalMFstate}), 
\begin{gather}
    B_{\textrm{CSL}}^{\pi_0}=\frac{16\pi m_{\pi}f_{\pi}^2}{\mu_{\textrm{B}}} \label{Bcsl_qcd} \,.
\end{gather}
On the other hand, as pointed out in ref.~\cite{Brauner:2016pko},
the charged pion fluctuations around the CSL state are tachyonic when 
the magnetic field strength is sufficiently large, $B>B_{\textrm{BEC}}^{\pi_{\pm}}$, with 
\begin{gather}
    B_{\textrm{BEC}}^{\pi_{\pm}}=\frac{m_{\pi}^2}{k^2}\left(2-k^2+2\sqrt{1-k^2+k^4} \right) \label{Bbec_qcd} \,,
\end{gather}
where $k$ is the elliptic modulus satisfying the following condition
(see eq.~(\ref{minimization_condition})):
\begin{gather}
    \frac{E(k)}{k}=\frac{\mu_{\textrm{B}}B_{\textrm{BEC}}^{\pi_{\pm}}}{16\pi m_{\pi}f_{\pi}^2} \label{kBEC_condition} \,.
\end{gather}
Eq.~(\ref{kBEC_condition}) determines 
$k$ as a function of
$\mu_{\textrm{B}}$ and $B_{\textrm{BEC}}^{\pi_{\pm}}$. 
Solving eq.~(\ref{Bbec_qcd}) with this $k$, we can determine the $\mu$ dependence of $B_{\textrm{BEC}}^{\pi_{\pm}}$.
It has been shown that $B_\mathrm{CSL}^{\pi_0} < B_\mathrm{BEC}^{\pi_\pm}$ 
for the parameter space of the interest~\cite{Brauner:2016pko}, 
and hence the neutral pion CSL can exist stably for the magnetic fields in this range.

The question is now whether the CSL can form quickly via nucleation in this range of the magnetic 
fields.
The critical magnetic field for the single wall disk formation (\ref{Bnucl}) can be applied into this case by the replacement $m\to m_{\pi}$ and $f \to f_{\pi}$:
\begin{gather}
    B_{\textrm{c}}^{\pi_0} = \frac{16\pi m_{\pi}f_{\pi}^2}{\mu_{\textrm{B}}}
    \left|\frac{(\ln (R_\mathrm{c}/r_\mathrm{c}))^{3/2}}{\sqrt{3}}\frac{4\pi^2f_{\pi}}{m_{\pi}}+1 \right| \label{Bnucl_qcd} \,.
\end{gather}
Figure~\ref{fig:critical_mag_qcd} show these critical magnetic field strength 
as a function of $\mu_B$, together with the lines $B=1 \mathrm{GeV}^2$ and $\mu_B=1 \mathrm{GeV}$, 
which indicates the upper bound of the validity of the effective theory described by pions~\cite{Shushpanov:1997sf,Agasian:1999sx}.
Here we take the physical values $f_{\pi} \approx 93\, \textrm{MeV}$ and $m_{\pi} \approx 140\, \textrm{MeV}$.
We can see that $B_c^{\pi_0} > B_\mathrm{BEC}^{\pi_\pm}$ is always satisfied. 
Moreover, in the parameter space with $\mu_B<1 \mathrm{GeV}$ where the pion effective theory is valid, 
$B_\mathrm{BEC}^{\pi_\pm} > B_\mathrm{CSL}^{\pi_0} $
is always satisfied. 
This suggests that for the parameter space of the interest, $B_\mathrm{CSL}^{\pi_0} < B<B_\mathrm{BEC}^{\pi_\pm}$ with $B<1 \mathrm{GeV}^2$ and $\mu_B<1 \mathrm{GeV}$,
even the single wall disk nucleation rate, and consequently that 
for the CSL, is exponentially suppressed, 
as long as our Nambu-Goto like effective description gives a good estimate for the nucleation rate.
We conclude that in order for the CSL to form quickly in the QCD system, 
we need a mechanism 
other than the topological defect formation through the quantum tunneling.

\subsection{Axion Like Particle} \label{sec:application_alp}

Next we consider an axon like particle (ALP) denoted as $a$~\cite{Preskill:1982cy,Abbott:1982af,Dine:1982ah,Svrcek:2006yi,Arvanitaki:2009fg,Marsh:2015xka}, 
which is a pseudo Nambu-Goldstone boson associated with a global U(1) symmetry breaking 
and is often predicted in the low-energy effective field theory of the string compactifications~\cite{Svrcek:2006yi,Conlon:2006tq,Arvanitaki:2009fg,Acharya:2010zx,Higaki:2011me,Cicoli:2012sz}.
Hence, ALPs are considered not only to be candidates of dark sector but also to show imprints of quantum gravity.
It is characterized by the mass  $m_a$, and axion decay constant $f_a$, which is also described by the effective potential
\begin{equation}
V(a) = m_a^2 f_a^2 \left(1 - \cos \frac{a}{f_a}\right), 
\end{equation}
with a canonical kinetic term, and the effective interactions to the SM sector, 
\begin{equation}
{\cal L} \ni - \frac{g_Y^2}{32\pi^2} C_Y \frac{a}{f_a} Y_{\mu\nu} {\tilde Y}^{\mu\nu} + \sum_i \frac{X_{iR}}{f_a} (\partial_\mu a) \psi_{i R}^\dagger \sigma^\mu \psi_{iR} - \sum_j \frac{X_{jL}}{f_a} (\partial_\mu a) \psi_{j L}^\dagger {\bar \sigma}^\mu \psi_{jL}, 
\end{equation}
where $g_Y$ is the hypergauge coupling constant, $Y_{\mu\nu}$ is the 
hypercharge gauge field strength, and $\psi_{iR}$ and $\psi_{jL}$ are the right- and left-handed Weyl fermions,
respectively.\footnote{Here we do not take into account the electroweak symmetry breaking, 
but the same discussion applies to the system with the spontaneously broken electroweak symmetry.} 
The coefficients $C_Y$ and $X_{iR/L}$ are determined by the UV physics. 
Note that the last two terms are the axion-fermion current coupling with 
\begin{equation}
j^\mu_{iR} \equiv \psi_{i R}^\dagger \sigma^\mu \psi_{iR} , \quad j^\mu_{jL} \equiv  \psi_{j L}^\dagger {\bar \sigma}^\mu \psi_{jL}. 
\end{equation}
Let us consider the case where 
the system has a uniform background hypermagnetic field ${\bm B}_Y$ and the chemical potentials for
each fermions, $\mu_{iR/L}$. 
In this case, the chiral magnetic effect induces the fermion current, 
\begin{equation}
{\bm j}_{i R/L} = (-1)^{\lambda_{iR/L}} \frac{\mu_{iR/L}}{2 \pi^2} (q_i g_Y) {\bm B}_Y, 
\end{equation}
where $\lambda_{iR}=0$ and $\lambda_{iL}=1$, respectively,  and $q_i$ is the hypercharge of the fermion $\psi_i$, 
which leads to the topological term, 
\begin{equation}
\mathcal{L}_\mathrm{eff} =\frac{g_Y}{2\pi^2} \left(\sum_i X_{iR} q_i \mu_{iR}   + \sum_j X_{jL} q_j \mu_{jL}   \right) {\bm B}_Y \cdot {\bm \nabla} \left(\frac{a}{f_a} \right).  
\end{equation}
Then with a similar discussion in Sec.~\ref{sec:nuck_CSL}, 
we can see that the axion CSL
also forms, 
whose properties can be read off with the replacement, 
\begin{equation}
\phi \rightarrow \frac{a}{f_a}, \quad  {\bm B} \rightarrow  {\bm B}_Y, \quad  \mu \rightarrow \mu_V \equiv 2g_Y  \left(\sum_i X_{iR} q_i \mu_{iR}   + \sum_j X_{jL} q_j \mu_{jL}   \right). 
\end{equation}
Consequently, the same properties to the CSL formed by neutral pion in the QCD,
including the formation rate, hold\footnote{
We have supposed that $a/f_a \to a/f_a + \theta$ under $U(1)_{\rm PQ}$ by an appropriate choice of chiral fermion charge.
}. 

The differences to the case of the CSL formed by pion in QCD 
are, i) we can freely take the axion mass and decay constant, 
ii) we do not have to worry about the tachyonic instability due to the charged pion fluctuation, 
and iii) the cutoff scale of the theory would be the Planck scale $M_\mathrm{pl}$ 
or the string scale $M_\mathrm{str}$. 
In the case $m_a \ll f_a \ll M_\mathrm{str}$, 
which is often considered
in conventional models or Large Volume Scenario \cite{Balasubramanian:2005zx,Conlon:2005ki,Conlon:2006tq,Cicoli:2012sz},
the critical magnetic field strength for the single domain wall formation 
in eq.~(\ref{Bnucl}) leads
\begin{equation}
B_{cY} \simeq \frac{64 \pi^2  f_a^3}{\mu_V}. \label{axionBc}
\end{equation}
Then we identify that for sufficiently large $\mu_V$, the critical magnetic field strength is well below 
the cutoff scale squared so that it seems that the CSL can easily form. 
However, this gives 
$B_{cY} \simeq \frac{f_a}{m_a}B_\mathrm{DW} \gg B_\mathrm{DW} = B_\mathrm{CSL}$, 
which makes the thin-wall approximation worse. See eqs.~\eqref{critical-mag} and~\eqref{validity_of_NGaction2}. 
On the other hand, if we consider a heavy ALP and take $m_a \lesssim f_a \ll M_\mathrm{str}$, 
the critical magnetic field strength is still approximated by eq.~\eqref{axionBc}, with being smaller than 
the cutoff scale, while not much larger than the critical magnetic field for the CSL
$B_{cY} \gtrsim B_\mathrm{CSL}$.
In this case, the thin-wall approximation is relatively good. 

One might wonder that the ALP CSL
can form in cosmology. 
Noting that at high temperature, $T > 10^5$ GeV, when several Yukawa interactions becomes ineffective, 
there are many approximate conserved charges within the SM~\cite{Campbell:1992jd,Garbrecht:2014kda,Domcke:2020kcp,Domcke:2020quw}, 
one can in principle have large chemical potential, $\mu_V \simeq T$, 
without suffering from the baryon overproduction.  
Then with our estimate in the above, we might expect the axion CSL
formation at 
$T \simeq f_a \gtrsim m_a \gg 10^5$ GeV if we have strong magnetic fields $B \simeq T^2$. 
However, the formation rate for the CSL
estimated in the above is the one at zero temperature. 
In order to investigate the possibility of the formation of the axion CSL
in the cosmic history, 
we need to develop its thermal formation rate, which is left for future study. 

\section{Conclusion and Discussion} \label{sec:Dis_Concl}
The ground state of the sine-Gordon theory 
with the background fermion chemical potential and magnetic fields, whose effect is implemented by
the total derivative term (\ref{original lagrangian}) 
is not the vacuum state ($\phi=0$), but the CSL state, 
for sufficiently large magnetic field, $B>B_{\textrm{CSL}}$
\cite{Brauner:2016pko}.
However, it has not been clear 
how the vacuum state changes to the CSL one.
In this work, we tackled this problem for the first time, in the best of our knowledge, by adopting
the Nambu-Goto like effective action for the 
string-domain wall disk system in 
the sine-Gordon theory with the topological term (\ref{original lagrangian}). 
The transition of the state from the vacuum to the CSL is described by the quantum mechanical tunneling
of the radius of a domain-wall disk surrounded by a string loop from $R_\mathrm{d}=0$ to $R_\mathrm{d} \not =0$, 
which is a similar approach to the studies in refs.~\cite{Basu:1991ig,Ai:2020vhx,Hayashi:2021kro}. 
In this description, for $B>B_{\textrm{CSL}}$ the wall tension becomes negative due to the topological term.  
Since the effective potential 
of the wall-disk radius $R$ has the potential barrier between $R=0$ and $R\rightarrow \infty$, 
because of the combination of the negative wall tension 
and the positive string tension, 
a domain wall-disk with a finite radius can be formed not classically but through the quantum tunneling.
In Section \ref{sec:cal_decay_rate},
we evaluated the nucleation rate of a 
single wall disk from the vacuum state as well as from the environment where the CSL has almost been formed, 
in terms of the bounce action eq.~\eqref{analytical_form_bounce_action} , which is the main result of the present paper. 
We regard that the latter represents the rate for the complete formation of the CSL. 
Due to
the repulsive force acting between the walls, which reduces the absolute value of the wall tension, the bounce action for the 
disk nucleation in the environment where the CSL has almost been formed is found to be larger than that from the vacuum, 
and hence the nucleation rate is more suppressed.
With these estimates, we determined the critical magnetic field for the disk nucleation 
where the exponential suppression becomes absent, ${\cal B} \simeq {\cal O}(1)$. 
Note that the thin-wall approximation is turned out likely to be violated for the disk with the critical radius for such large magnetic field strength. 
However, we expect that it gives a good estimate for the disk nucleation from the discussion in ref.~\cite{Hayashi:2021kro}, 
where the exponential suppression factor for the larger radius is turned to be equal to that for the critical radius in the case of the three-dimensional bubble nucleation. 
There it has been identified as the whole quantum estimate of the rate for the quantum tunneling followed by the classical expansion.

Since our analysis is based on the sine-Gordon theory with the total derivative term,
the results in section \ref{sec:cal_decay_rate} can be applied to systems described by the same Lagrangian as eq.~(\ref{original lagrangian}).
In section \ref{sec:application_qcd}, we apply our analysis into 2-flavor QCD at finite baryon chemical potential under an external magnetic field.
We found that within the scope of the low energy effective theory, the formation of 
a single domain-wall disk from the vacuum is exponentially suppressed, $\propto e^{-{\cal B}}, \ {\cal B} \gg 1$.
Therefore, even when the CSL is more stable than the vacuum state at $B>B_{\textrm{CSL}}$, the vacuum state cannot 
be transformed into the CSL one quickly.
However, there may be loopholes in the above discussion.
As we have explained in the above, the thin-wall approximation for the validity of the Nambu-Goto action 
is not a good approximation for the disk with a critical radius, $R_2$.
For example, substituting $f_{\pi} \approx 93\, \textrm{MeV}$, $m_{\pi} \approx 140\, \textrm{MeV}$ and $B=2B_{\textrm{DW}}$ for eqs.~(\ref{validity_of_NGaction1}) and \eqref{validity_of_NGaction2},
the ratios of the vortex size $r_\mathrm{c}$ as well as the wall thickness $r_\mathrm{w}$ 
to critical disk radius 
$R_\mathrm{d} = R_2$ are evaluated as
\begin{gather}
    \frac{r_{\textrm{c}}}{R_2} \sim 0.5 \,, \quad  \frac{r_{\textrm{w}}}{R_2} \sim 0.3 \,,
\end{gather}
where we have approximated the logarithmic factor to be unity. Apparently,
it is not sufficiently smaller than $1$.
Therefore, the Nambu-Goto type effective action may not be applicable in the CSL of QCD.
Note that this arguments also follows for the ALP CSL.
The other potential loophole is the effects of the charged pions.
We have considered the quantum nucleation of the topological soliton with the topological number $\pi_1(\textrm{U}(1)) \simeq \mathbb{Z}$ so far.
Although the charged pions acquire masses by the Landau quantization, sufficiently heavy to be neglected,
the actual configuration space of the mesons ($\pi_0$, $\pi_{\pm}$) is $\textrm{SU}(2)$ instead of $\textrm{U}(1)$.
Hence, 
while we expect that the charged pions do not show nontrivial field configuration during the disk formation process
due to their heavy mass, we do not exclude the possibility that the wall disk is formed
continuously from the QCD vacuum due to $\pi_1(\textrm{SU}(2))\simeq 0$ when we include the degrees of freedom of the charged mesons.
We leave more detailed investigation in future study. 
Note that in the case of ALPs, which is studied in Sec.~\ref{sec:application_alp}, 
we have considered a single ALP and it is assumed that there exist no associated light charged field in the theory.
If this is the case in the string theory, loopholes would not apply.
However, in general there may have to exist associated fields, e.g., lots of ALPs, in string theory 
owing to the swampland constraints restricting the moduli space \cite{Vafa:2005ui,Palti:2019pca}.

\

\noindent 
{\it Note added}: While completing this paper, we learned another study on the formation of the chiral soliton lattice~\cite{Eto:2022lhu}
by M.~Eto and M.~Nitta. 

\section*{Acknowledgements}
The authors would like to thank M.~Eto and M.~Nitta for useful comments and for informing us of their study on the soliton formation.
The authors would also like to thank M. Hongo for discussion.
This work is supported by JSPS KAKENHI, Grant-in-Aid for Scientific Research (C) 22K03601 (T.\,H.) and 
JP19K03842 (K.\,K.). 
K.~N. is supported by JSPS KAKENHI, Grant-in-Aid for Scientific Research (B) 21H01084.
\bibliographystyle{utphys}
\bibliography{reference.bib}

\providecommand{\href}[2]{#2}\begingroup\raggedright\begin{thebibliography}{10}

\bibitem{dzyaloshinskii1964theory}
I.~Dzyaloshinskii, ``Theory of helicoidal structures in antiferromagnets. i.
  nonmetals,'' {\em Sov. Phys. JETP} {\bfseries 19} no.~4, (1964) 960--971.

\bibitem{togawa2016symmetry}
Y.~Togawa, Y.~Kousaka, K.~Inoue, and J.-i. Kishine, ``Symmetry, structure, and
  dynamics of monoaxial chiral magnets,'' {\em Journal of the Physical Society
  of Japan} {\bfseries 85} no.~11, (2016) 112001.

\bibitem{kishine2015theory}
J.-i. Kishine and A.~Ovchinnikov, ``Theory of monoaxial chiral helimagnet,''
  {\em Solid State Physics} {\bfseries 66} (2015) 1--130.

\bibitem{Son:2004tq}
D.~T. Son and A.~R. Zhitnitsky, ``{Quantum anomalies in dense matter},''
  \href{http://dx.doi.org/10.1103/PhysRevD.70.074018}{{\em Phys. Rev. D}
  {\bfseries 70} (2004) 074018},
  \href{http://arxiv.org/abs/hep-ph/0405216}{{\ttfamily arXiv:hep-ph/0405216}}.

\bibitem{Son:2007ny}
D.~T. Son and M.~A. Stephanov, ``{Axial anomaly and magnetism of nuclear and
  quark matter},'' \href{http://dx.doi.org/10.1103/PhysRevD.77.014021}{{\em
  Phys. Rev. D} {\bfseries 77} (2008) 014021},
  \href{http://arxiv.org/abs/0710.1084}{{\ttfamily arXiv:0710.1084 [hep-ph]}}.

\bibitem{Brauner:2016pko}
T.~Brauner and N.~Yamamoto, ``{Chiral Soliton Lattice and Charged Pion
  Condensation in Strong Magnetic Fields},''
  \href{http://dx.doi.org/10.1007/JHEP04(2017)132}{{\em JHEP} {\bfseries 04}
  (2017) 132}, \href{http://arxiv.org/abs/1609.05213}{{\ttfamily
  arXiv:1609.05213 [hep-ph]}}.

\bibitem{Eto:2012qd}
M.~Eto, K.~Hashimoto, and T.~Hatsuda, ``{Ferromagnetic neutron stars: axial
  anomaly, dense neutron matter, and pionic wall},''
  \href{http://dx.doi.org/10.1103/PhysRevD.88.081701}{{\em Phys. Rev. D}
  {\bfseries 88} (2013) 081701},
  \href{http://arxiv.org/abs/1209.4814}{{\ttfamily arXiv:1209.4814 [hep-ph]}}.

\bibitem{Kawaguchi:2018fpi}
M.~Kawaguchi, Y.-L. Ma, and S.~Matsuzaki, ``{Chiral soliton lattice effect on
  baryonic matter from a skyrmion crystal model},''
  \href{http://dx.doi.org/10.1103/PhysRevC.100.025207}{{\em Phys. Rev. C}
  {\bfseries 100} no.~2, (2019) 025207},
  \href{http://arxiv.org/abs/1810.12880}{{\ttfamily arXiv:1810.12880
  [nucl-th]}}.

\bibitem{Chen:2021vou}
S.~Chen, K.~Fukushima, and Z.~Qiu, ``{Skyrmions in a magnetic field and
  \ensuremath{\pi}0 domain wall formation in dense nuclear matter},''
  \href{http://dx.doi.org/10.1103/PhysRevD.105.L011502}{{\em Phys. Rev. D}
  {\bfseries 105} no.~1, (2022) L011502},
  \href{http://arxiv.org/abs/2104.11482}{{\ttfamily arXiv:2104.11482
  [hep-ph]}}.

\bibitem{Gronli:2022cri}
M.~S. Gr\o{}nli and T.~Brauner, ``{Competition of chiral soliton lattice and
  Abrikosov vortex lattice in QCD with isospin chemical potential},''
  \href{http://dx.doi.org/10.1140/epjc/s10052-022-10300-5}{{\em Eur. Phys. J.
  C} {\bfseries 82} no.~4, (2022) 354},
  \href{http://arxiv.org/abs/2201.07065}{{\ttfamily arXiv:2201.07065
  [hep-ph]}}.

\bibitem{Yamamoto:2015maz}
N.~Yamamoto, ``{Axion electrodynamics and nonrelativistic photons in nuclear
  and quark matter},'' \href{http://dx.doi.org/10.1103/PhysRevD.93.085036}{{\em
  Phys. Rev. D} {\bfseries 93} no.~8, (2016) 085036},
  \href{http://arxiv.org/abs/1512.05668}{{\ttfamily arXiv:1512.05668
  [hep-th]}}.

\bibitem{Brauner:2017mui}
T.~Brauner and S.~Kadam, ``{Anomalous electrodynamics of neutral pion matter in
  strong magnetic fields},''
  \href{http://dx.doi.org/10.1007/JHEP03(2017)015}{{\em JHEP} {\bfseries 03}
  (2017) 015}, \href{http://arxiv.org/abs/1701.06793}{{\ttfamily
  arXiv:1701.06793 [hep-ph]}}.

\bibitem{Brauner:2021sci}
T.~Brauner, H.~Kole\v{s}ov\'a, and N.~Yamamoto, ``{Chiral soliton lattice phase
  in warm QCD},'' \href{http://dx.doi.org/10.1016/j.physletb.2021.136767}{{\em
  Phys. Lett. B} {\bfseries 823} (2021) 136767},
  \href{http://arxiv.org/abs/2108.10044}{{\ttfamily arXiv:2108.10044
  [hep-ph]}}.

\bibitem{Evans:2022hwr}
G.~W. Evans and A.~Schmitt, ``{Chiral anomaly induces superconducting baryon
  crystal},'' \href{http://arxiv.org/abs/2206.01227}{{\ttfamily
  arXiv:2206.01227 [hep-th]}}.

\bibitem{Huang:2017pqe}
X.-G. Huang, K.~Nishimura, and N.~Yamamoto, ``{Anomalous effects of dense
  matter under rotation},''
  \href{http://dx.doi.org/10.1007/JHEP02(2018)069}{{\em JHEP} {\bfseries 02}
  (2018) 069}, \href{http://arxiv.org/abs/1711.02190}{{\ttfamily
  arXiv:1711.02190 [hep-ph]}}.

\bibitem{Nishimura:2020odq}
K.~Nishimura and N.~Yamamoto, ``{Topological term, QCD anomaly, and the
  $\eta^{'}$ chiral soliton lattice in rotating baryonic matter},''
  \href{http://dx.doi.org/10.1007/JHEP07(2020)196}{{\em JHEP} {\bfseries 07}
  no.~07, (2020) 196}, \href{http://arxiv.org/abs/2003.13945}{{\ttfamily
  arXiv:2003.13945 [hep-ph]}}.

\bibitem{Eto:2021gyy}
M.~Eto, K.~Nishimura, and M.~Nitta, ``{Phases of rotating baryonic matter:
  non-Abelian chiral soliton lattices, antiferro-isospin chains, and
  ferri/ferromagnetic magnetization},''
  \href{http://arxiv.org/abs/2112.01381}{{\ttfamily arXiv:2112.01381
  [hep-ph]}}.

\bibitem{Chen:2021aiq}
H.-L. Chen, X.-G. Huang, and J.~Liao, ``{QCD phase structure under rotation},''
  \href{http://dx.doi.org/10.1007/978-3-030-71427-7_11}{{\em Lect. Notes Phys.}
  {\bfseries 987} (2021) 349--379},
  \href{http://arxiv.org/abs/2108.00586}{{\ttfamily arXiv:2108.00586
  [hep-ph]}}.

\bibitem{Yamada:2021jhy}
A.~Yamada and N.~Yamamoto, ``{Floquet vacuum engineering: Laser-driven chiral
  soliton lattice in the QCD vacuum},''
  \href{http://dx.doi.org/10.1103/PhysRevD.104.054041}{{\em Phys. Rev. D}
  {\bfseries 104} no.~5, (2021) 054041},
  \href{http://arxiv.org/abs/2107.07074}{{\ttfamily arXiv:2107.07074
  [hep-ph]}}.

\bibitem{Brauner:2019aid}
T.~Brauner, G.~Filios, and H.~Kole\v{s}ov\'a, ``{Chiral soliton lattice in
  QCD-like theories},'' \href{http://dx.doi.org/10.1007/JHEP12(2019)029}{{\em
  JHEP} {\bfseries 12} (2019) 029},
  \href{http://arxiv.org/abs/1905.11409}{{\ttfamily arXiv:1905.11409
  [hep-ph]}}.

\bibitem{Brauner:2019rjg}
T.~Brauner, G.~Filios, and H.~Kole\v{s}ov\'a, ``{Anomaly-Induced Inhomogeneous
  Phase in Quark Matter without the Sign Problem},''
  \href{http://dx.doi.org/10.1103/PhysRevLett.123.012001}{{\em Phys. Rev.
  Lett.} {\bfseries 123} no.~1, (2019) 012001},
  \href{http://arxiv.org/abs/1902.07522}{{\ttfamily arXiv:1902.07522
  [hep-ph]}}.

\bibitem{Yamamoto:2018hdy}
N.~Yamamoto, ``{Chirality Driven Helical Pattern Formation},''
  \href{http://arxiv.org/abs/1808.00326}{{\ttfamily arXiv:1808.00326
  [cond-mat.stat-mech]}}.

\bibitem{Vilenkin:1980fu}
A.~Vilenkin, ``{EQUILIBRIUM PARITY VIOLATING CURRENT IN A MAGNETIC FIELD},''
  \href{http://dx.doi.org/10.1103/PhysRevD.22.3080}{{\em Phys. Rev. D}
  {\bfseries 22} (1980) 3080--3084}.

\bibitem{Metlitski:2005pr}
M.~A. Metlitski and A.~R. Zhitnitsky, ``{Anomalous axion interactions and
  topological currents in dense matter},''
  \href{http://dx.doi.org/10.1103/PhysRevD.72.045011}{{\em Phys. Rev. D}
  {\bfseries 72} (2005) 045011},
  \href{http://arxiv.org/abs/hep-ph/0505072}{{\ttfamily arXiv:hep-ph/0505072}}.

\bibitem{Fukushima:2008xe}
K.~Fukushima, D.~E. Kharzeev, and H.~J. Warringa, ``{The Chiral Magnetic
  Effect},'' \href{http://dx.doi.org/10.1103/PhysRevD.78.074033}{{\em Phys.
  Rev. D} {\bfseries 78} (2008) 074033},
  \href{http://arxiv.org/abs/0808.3382}{{\ttfamily arXiv:0808.3382 [hep-ph]}}.

\bibitem{Vilenkin:1979ui}
A.~Vilenkin, ``{MACROSCOPIC PARITY VIOLATING EFFECTS: NEUTRINO FLUXES FROM
  ROTATING BLACK HOLES AND IN ROTATING THERMAL RADIATION},''
  \href{http://dx.doi.org/10.1103/PhysRevD.20.1807}{{\em Phys. Rev. D}
  {\bfseries 20} (1979) 1807--1812}.

\bibitem{Vilenkin:1980zv}
A.~Vilenkin, ``{QUANTUM FIELD THEORY AT FINITE TEMPERATURE IN A ROTATING
  SYSTEM},'' \href{http://dx.doi.org/10.1103/PhysRevD.21.2260}{{\em Phys. Rev.
  D} {\bfseries 21} (1980) 2260--2269}.

\bibitem{Landsteiner:2011cp}
K.~Landsteiner, E.~Megias, and F.~Pena-Benitez, ``{Gravitational Anomaly and
  Transport},'' \href{http://dx.doi.org/10.1103/PhysRevLett.107.021601}{{\em
  Phys. Rev. Lett.} {\bfseries 107} (2011) 021601},
  \href{http://arxiv.org/abs/1103.5006}{{\ttfamily arXiv:1103.5006 [hep-ph]}}.

\bibitem{Landsteiner:2012kd}
K.~Landsteiner, E.~Meg\'{i}as, and F.~Pena-Benitez, ``{Anomalous Transport from
  Kubo Formulae},'' \href{http://dx.doi.org/10.1007/978-3-642-37305-3_17}{{\em
  Lect. Notes Phys.} {\bfseries 871} (2013) 433--468},
\href{http://arxiv.org/abs/1207.5808}{{\ttfamily arXiv:1207.5808 [hep-th]}}.

\bibitem{Landsteiner:2016led}
K.~Landsteiner, ``{Notes on Anomaly Induced Transport},''
  \href{http://dx.doi.org/10.5506/APhysPolB.47.2617}{{\em Acta Phys. Polon.}
  {\bfseries B47} (2016) 2617},
\href{http://arxiv.org/abs/1610.04413}{{\ttfamily arXiv:1610.04413 [hep-th]}}.

\bibitem{Son:2009tf}
D.~T. Son and P.~Surowka, ``{Hydrodynamics with Triangle Anomalies},''
  \href{http://dx.doi.org/10.1103/PhysRevLett.103.191601}{{\em Phys. Rev.
  Lett.} {\bfseries 103} (2009) 191601},
  \href{http://arxiv.org/abs/0906.5044}{{\ttfamily arXiv:0906.5044 [hep-th]}}.

\bibitem{DZYALOSHINSKY1958241}
I.~Dzyaloshinsky, ``A thermodynamic theory of “weak” ferromagnetism of
  antiferromagnetics,''
  \href{http://dx.doi.org/https://doi.org/10.1016/0022-3697(58)90076-3}{{\em
  Journal of Physics and Chemistry of Solids} {\bfseries 4} no.~4, (1958)
  241--255}.
  \url{https://www.sciencedirect.com/science/article/pii/0022369758900763}.

\bibitem{PhysRev.120.91}
T.~Moriya, ``Anisotropic superexchange interaction and weak ferromagnetism,''
  \href{http://dx.doi.org/10.1103/PhysRev.120.91}{{\em Phys. Rev.} {\bfseries
  120} (Oct, 1960) 91--98}.
  \url{https://link.aps.org/doi/10.1103/PhysRev.120.91}.

\bibitem{Kibble:1976sj}
T.~W.~B. Kibble, ``{Topology of Cosmic Domains and Strings},''
  \href{http://dx.doi.org/10.1088/0305-4470/9/8/029}{{\em J. Phys. A}
  {\bfseries 9} (1976) 1387--1398}.

\bibitem{Zurek:1985qw}
W.~H. Zurek, ``{Cosmological Experiments in Superfluid Helium?},''
  \href{http://dx.doi.org/10.1038/317505a0}{{\em Nature} {\bfseries 317} (1985)
  505--508}.

\bibitem{paterson2019order}
G.~W. Paterson, T.~Koyama, M.~Shinozaki, Y.~Masaki, F.~J. Goncalves,
  Y.~Shimamoto, T.~Sogo, M.~Nord, Y.~Kousaka, Y.~Kato, {\em et~al.}, ``Order
  and disorder in the magnetization of the chiral crystal crnb 3 s 6,''
  \href{http://dx.doi.org/https://doi.org/10.1103/PhysRevB.99.224429}{{\em
  Physical Review B} {\bfseries 99} no.~22, (2019) 224429}.

\bibitem{Basu:1991ig}
R.~Basu, A.~H. Guth, and A.~Vilenkin, ``{Quantum creation of topological
  defects during inflation},''
  \href{http://dx.doi.org/10.1103/PhysRevD.44.340}{{\em Phys. Rev. D}
  {\bfseries 44} (1991) 340--351}.

\bibitem{Ai:2020vhx}
W.-Y. Ai and M.~Drewes, ``{Schwinger effect and false vacuum decay as
  quantum-mechanical tunneling of a relativistic particle},''
  \href{http://dx.doi.org/10.1103/PhysRevD.102.076015}{{\em Phys. Rev. D}
  {\bfseries 102} no.~7, (2020) 076015},
  \href{http://arxiv.org/abs/2005.14163}{{\ttfamily arXiv:2005.14163
  [hep-th]}}.

\bibitem{Hayashi:2021kro}
T.~Hayashi, K.~Kamada, N.~Oshita, and J.~Yokoyama, ``{Vacuum decay in the
  Lorentzian path integral},''
  \href{http://dx.doi.org/10.1088/1475-7516/2022/05/041}{{\em JCAP} {\bfseries
  05} no.~05, (2022) 041}, \href{http://arxiv.org/abs/2112.09284}{{\ttfamily
  arXiv:2112.09284 [hep-th]}}.

\bibitem{Svrcek:2006yi}
P.~Svrcek and E.~Witten, ``{Axions In String Theory},''
  \href{http://dx.doi.org/10.1088/1126-6708/2006/06/051}{{\em JHEP} {\bfseries
  06} (2006) 051}, \href{http://arxiv.org/abs/hep-th/0605206}{{\ttfamily
  arXiv:hep-th/0605206}}.

\bibitem{Conlon:2006tq}
J.~P. Conlon, ``{The QCD axion and moduli stabilisation},''
  \href{http://dx.doi.org/10.1088/1126-6708/2006/05/078}{{\em JHEP} {\bfseries
  05} (2006) 078}, \href{http://arxiv.org/abs/hep-th/0602233}{{\ttfamily
  arXiv:hep-th/0602233}}.

\bibitem{Arvanitaki:2009fg}
A.~Arvanitaki, S.~Dimopoulos, S.~Dubovsky, N.~Kaloper, and J.~March-Russell,
  ``{String Axiverse},''
  \href{http://dx.doi.org/10.1103/PhysRevD.81.123530}{{\em Phys. Rev. D}
  {\bfseries 81} (2010) 123530},
  \href{http://arxiv.org/abs/0905.4720}{{\ttfamily arXiv:0905.4720 [hep-th]}}.

\bibitem{Acharya:2010zx}
B.~S. Acharya, K.~Bobkov, and P.~Kumar, ``{An M Theory Solution to the Strong
  CP Problem and Constraints on the Axiverse},''
  \href{http://dx.doi.org/10.1007/JHEP11(2010)105}{{\em JHEP} {\bfseries 11}
  (2010) 105}, \href{http://arxiv.org/abs/1004.5138}{{\ttfamily arXiv:1004.5138
  [hep-th]}}.

\bibitem{Higaki:2011me}
T.~Higaki and T.~Kobayashi, ``{Note on moduli stabilization, supersymmetry
  breaking and axiverse},''
  \href{http://dx.doi.org/10.1103/PhysRevD.84.045021}{{\em Phys. Rev. D}
  {\bfseries 84} (2011) 045021},
  \href{http://arxiv.org/abs/1106.1293}{{\ttfamily arXiv:1106.1293 [hep-th]}}.

\bibitem{Cicoli:2012sz}
M.~Cicoli, M.~Goodsell, and A.~Ringwald, ``{The type IIB string axiverse and
  its low-energy phenomenology},''
  \href{http://dx.doi.org/10.1007/JHEP10(2012)146}{{\em JHEP} {\bfseries 10}
  (2012) 146}, \href{http://arxiv.org/abs/1206.0819}{{\ttfamily arXiv:1206.0819
  [hep-th]}}.

\bibitem{Marsh:2015xka}
D.~J.~E. Marsh, ``{Axion Cosmology},''
  \href{http://dx.doi.org/10.1016/j.physrep.2016.06.005}{{\em Phys. Rept.}
  {\bfseries 643} (2016) 1--79},
  \href{http://arxiv.org/abs/1510.07633}{{\ttfamily arXiv:1510.07633
  [astro-ph.CO]}}.

\bibitem{Son:2012wh}
D.~T. Son and N.~Yamamoto, ``{Berry Curvature, Triangle Anomalies, and the
  Chiral Magnetic Effect in Fermi Liquids},''
  \href{http://dx.doi.org/10.1103/PhysRevLett.109.181602}{{\em Phys. Rev.
  Lett.} {\bfseries 109} (2012) 181602},
  \href{http://arxiv.org/abs/1203.2697}{{\ttfamily arXiv:1203.2697
  [cond-mat.mes-hall]}}.

\bibitem{Fukushima:2018ohd}
K.~Fukushima and S.~Imaki, ``{Anomaly inflow on QCD axial domain-walls and
  vortices},'' \href{http://dx.doi.org/10.1103/PhysRevD.97.114003}{{\em Phys.
  Rev. D} {\bfseries 97} no.~11, (2018) 114003},
  \href{http://arxiv.org/abs/1802.08096}{{\ttfamily arXiv:1802.08096
  [hep-ph]}}.

\bibitem{Kaplan:2001hh}
D.~B. Kaplan and S.~Reddy, ``{Charged and superconducting vortices in dense
  quark matter},'' \href{http://dx.doi.org/10.1103/PhysRevLett.88.132302}{{\em
  Phys. Rev. Lett.} {\bfseries 88} (2002) 132302},
  \href{http://arxiv.org/abs/hep-ph/0109256}{{\ttfamily arXiv:hep-ph/0109256}}.

\bibitem{Buckley:2002ur}
K.~B.~W. Buckley and A.~R. Zhitnitsky, ``{Superconducting K strings in high
  density QCD},'' \href{http://dx.doi.org/10.1088/1126-6708/2002/08/013}{{\em
  JHEP} {\bfseries 08} (2002) 013},
  \href{http://arxiv.org/abs/hep-ph/0204064}{{\ttfamily arXiv:hep-ph/0204064}}.

\bibitem{Zeldovich:1974uw}
Y.~B. Zeldovich, I.~Y. Kobzarev, and L.~B. Okun, ``{Cosmological Consequences
  of the Spontaneous Breakdown of Discrete Symmetry},'' {\em Zh. Eksp. Teor.
  Fiz.} {\bfseries 67} (1974) 3--11.

\bibitem{Kibble:1982dd}
T.~W.~B. Kibble, G.~Lazarides, and Q.~Shafi, ``{Walls Bounded by Strings},''
  \href{http://dx.doi.org/10.1103/PhysRevD.26.435}{{\em Phys. Rev. D}
  {\bfseries 26} (1982) 435}.

\bibitem{Vilenkin:1982ks}
A.~Vilenkin and A.~E. Everett, ``{Cosmic Strings and Domain Walls in Models
  with Goldstone and PseudoGoldstone Bosons},''
  \href{http://dx.doi.org/10.1103/PhysRevLett.48.1867}{{\em Phys. Rev. Lett.}
  {\bfseries 48} (1982) 1867--1870}.

\bibitem{Vilenkin:2000jqa}
A.~Vilenkin and E.~P.~S. Shellard, {\em {Cosmic Strings and Other Topological
  Defects}}.
\newblock Cambridge University Press, 7, 2000.

\bibitem{Hindmarsh:1994re}
M.~B. Hindmarsh and T.~W.~B. Kibble, ``{Cosmic strings},''
  \href{http://dx.doi.org/10.1088/0034-4885/58/5/001}{{\em Rept. Prog. Phys.}
  {\bfseries 58} (1995) 477--562},
  \href{http://arxiv.org/abs/hep-ph/9411342}{{\ttfamily arXiv:hep-ph/9411342}}.

\bibitem{Chatterjee:2019rch}
C.~Chatterjee, T.~Higaki, and M.~Nitta, ``{Note on a solution to domain wall
  problem with the Lazarides-Shafi mechanism in axion dark matter models},''
  \href{http://dx.doi.org/10.1103/PhysRevD.101.075026}{{\em Phys. Rev. D}
  {\bfseries 101} no.~7, (2020) 075026},
  \href{http://arxiv.org/abs/1903.11753}{{\ttfamily arXiv:1903.11753
  [hep-ph]}}.

\bibitem{Eto:2022lhu}
M.~Eto and M.~Nitta, ``{Quantum nucleation of topological solitons},''
  \href{http://arxiv.org/abs/2207.00211}{{\ttfamily arXiv:2207.00211
  [hep-th]}}.

\bibitem{Coleman:1977py}
S.~R. Coleman, ``{The Fate of the False Vacuum. 1. Semiclassical Theory},''
  \href{http://dx.doi.org/10.1103/PhysRevD.16.1248}{{\em Phys. Rev. D}
  {\bfseries 15} (1977) 2929--2936}. [Erratum: Phys.Rev.D 16, 1248 (1977)].

\bibitem{Callan:1977pt}
C.~G. Callan, Jr. and S.~R. Coleman, ``{The Fate of the False Vacuum. 2. First
  Quantum Corrections},''
  \href{http://dx.doi.org/10.1103/PhysRevD.16.1762}{{\em Phys. Rev. D}
  {\bfseries 16} (1977) 1762--1768}.

\bibitem{Weinberg:2012pjx}
E.~J. Weinberg, \href{http://dx.doi.org/10.1017/CBO9781139017787}{{\em
  {Classical solutions in quantum field theory}: {Solitons and Instantons in
  High Energy Physics}}}.
\newblock Cambridge Monographs on Mathematical Physics. Cambridge University
  Press, 9, 2012.

\bibitem{Shushpanov:1997sf}
I.~A. Shushpanov and A.~V. Smilga, ``{Quark condensate in a magnetic field},''
  \href{http://dx.doi.org/10.1016/S0370-2693(97)00441-3}{{\em Phys. Lett. B}
  {\bfseries 402} (1997) 351--358},
  \href{http://arxiv.org/abs/hep-ph/9703201}{{\ttfamily arXiv:hep-ph/9703201}}.

\bibitem{Agasian:1999sx}
N.~O. Agasian and I.~A. Shushpanov, ``{The Quark and gluon condensates and
  low-energy QCD theorems in a magnetic field},''
  \href{http://dx.doi.org/10.1016/S0370-2693(99)01414-8}{{\em Phys. Lett. B}
  {\bfseries 472} (2000) 143--149},
  \href{http://arxiv.org/abs/hep-ph/9911254}{{\ttfamily arXiv:hep-ph/9911254}}.

\bibitem{Preskill:1982cy}
J.~Preskill, M.~B. Wise, and F.~Wilczek, ``{Cosmology of the Invisible
  Axion},'' \href{http://dx.doi.org/10.1016/0370-2693(83)90637-8}{{\em Phys.
  Lett. B} {\bfseries 120} (1983) 127--132}.

\bibitem{Abbott:1982af}
L.~F. Abbott and P.~Sikivie, ``{A Cosmological Bound on the Invisible Axion},''
  \href{http://dx.doi.org/10.1016/0370-2693(83)90638-X}{{\em Phys. Lett. B}
  {\bfseries 120} (1983) 133--136}.

\bibitem{Dine:1982ah}
M.~Dine and W.~Fischler, ``{The Not So Harmless Axion},''
  \href{http://dx.doi.org/10.1016/0370-2693(83)90639-1}{{\em Phys. Lett. B}
  {\bfseries 120} (1983) 137--141}.

\bibitem{Balasubramanian:2005zx}
V.~Balasubramanian, P.~Berglund, J.~P. Conlon, and F.~Quevedo, ``{Systematics
  of moduli stabilisation in Calabi-Yau flux compactifications},''
  \href{http://dx.doi.org/10.1088/1126-6708/2005/03/007}{{\em JHEP} {\bfseries
  03} (2005) 007}, \href{http://arxiv.org/abs/hep-th/0502058}{{\ttfamily
  arXiv:hep-th/0502058}}.

\bibitem{Conlon:2005ki}
J.~P. Conlon, F.~Quevedo, and K.~Suruliz, ``{Large-volume flux
  compactifications: Moduli spectrum and D3/D7 soft supersymmetry breaking},''
  \href{http://dx.doi.org/10.1088/1126-6708/2005/08/007}{{\em JHEP} {\bfseries
  08} (2005) 007}, \href{http://arxiv.org/abs/hep-th/0505076}{{\ttfamily
  arXiv:hep-th/0505076}}.

\bibitem{Campbell:1992jd}
B.~A. Campbell, S.~Davidson, J.~R. Ellis, and K.~A. Olive, ``{On the baryon,
  lepton flavor and right-handed electron asymmetries of the universe},''
  \href{http://dx.doi.org/10.1016/0370-2693(92)91079-O}{{\em Phys. Lett. B}
  {\bfseries 297} (1992) 118--124},
  \href{http://arxiv.org/abs/hep-ph/9302221}{{\ttfamily arXiv:hep-ph/9302221}}.

\bibitem{Garbrecht:2014kda}
B.~Garbrecht and P.~Schwaller, ``{Spectator Effects during Leptogenesis in the
  Strong Washout Regime},''
  \href{http://dx.doi.org/10.1088/1475-7516/2014/10/012}{{\em JCAP} {\bfseries
  10} (2014) 012}, \href{http://arxiv.org/abs/1404.2915}{{\ttfamily
  arXiv:1404.2915 [hep-ph]}}.

\bibitem{Domcke:2020kcp}
V.~Domcke, Y.~Ema, K.~Mukaida, and M.~Yamada, ``{Spontaneous Baryogenesis from
  Axions with Generic Couplings},''
  \href{http://dx.doi.org/10.1007/JHEP08(2020)096}{{\em JHEP} {\bfseries 08}
  (2020) 096}, \href{http://arxiv.org/abs/2006.03148}{{\ttfamily
  arXiv:2006.03148 [hep-ph]}}.

\bibitem{Domcke:2020quw}
V.~Domcke, K.~Kamada, K.~Mukaida, K.~Schmitz, and M.~Yamada, ``{Wash-In
  Leptogenesis},'' \href{http://dx.doi.org/10.1103/PhysRevLett.126.201802}{{\em
  Phys. Rev. Lett.} {\bfseries 126} no.~20, (2021) 201802},
  \href{http://arxiv.org/abs/2011.09347}{{\ttfamily arXiv:2011.09347
  [hep-ph]}}.

\bibitem{Vafa:2005ui}
C.~Vafa, ``{The String landscape and the swampland},''
  \href{http://arxiv.org/abs/hep-th/0509212}{{\ttfamily arXiv:hep-th/0509212}}.

\bibitem{Palti:2019pca}
E.~Palti, ``{The Swampland: Introduction and Review},''
  \href{http://dx.doi.org/10.1002/prop.201900037}{{\em Fortsch. Phys.}
  {\bfseries 67} no.~6, (2019) 1900037},
  \href{http://arxiv.org/abs/1903.06239}{{\ttfamily arXiv:1903.06239
  [hep-th]}}.

\end{thebibliography}\endgroup


\providecommand{\href}[2]{#2}\begingroup\raggedright\endgroup

\end{document}